\documentclass[aps, prd, twocolumn, amsmath, floats,floatfix, superscriptaddress, nofootinbib, showpacs]{revtex4}

\usepackage{braket}
\usepackage{siunitx}
\usepackage{graphicx}
\usepackage{epsfig}
\usepackage{amsmath,amsfonts,amssymb}
\usepackage{caption}
\usepackage{subcaption}
\usepackage{wrapfig}
\usepackage{diagbox}
\usepackage{nicefrac}
\usepackage{verbatim}
\usepackage{float}
\usepackage{morefloats}
\usepackage[dvipsnames]{xcolor}
\usepackage{booktabs}
\usepackage[normalem]{ulem}
\usepackage{multirow}
\usepackage{makecell}
\usepackage{tabulary}
\usepackage{cancel}
\usepackage{tikz}
\usetikzlibrary{positioning}
\usepackage{graphicx}
\usepackage{xcolor}

\usepackage[capitalise]{cleveref}

\usepackage{ragged2e}
\usepackage{subcaption}   
\DeclareCaptionJustification{justified}{\justifying}
\usepackage{comment}
\crefname{section}{Sec.}{Secs.}
\crefname{table}{Tab.}{Tabs.}
\crefname{figure}{Fig.}{Figs.}
\crefname{equation}{Eq.}{Eqs.}
\crefname{appendix}{Appendix\ }{Appendix\ }

\providecommand{\openone}{\leavevmode\hbox{\small1\kern-3.8pt\normalsize1}}

\DeclareSIUnit\parsec{pc}
\DeclareSIUnit\megaparsec{Mpc}
\DeclareSIUnit[quantity-product = {}]\solar{\text{$\mathrm{M}_\odot$}}

\definecolor{bostonuniversityred}{rgb}{0.8, 0.0, 0.0}

\usepackage{booktabs}
\usepackage[Q=yes,pverb-linebreak=no]{examplep}
\begin{document}

\title{Comparison of neural network architectures for feature extraction from binary black hole merger waveforms}

\author{Osvaldo~Gramaxo~Freitas}
\email{ogf1996@gmail.com}
\affiliation{Centro de F\'{\i}sica das Universidades do Minho e do Porto (CF-UM-UP), Universidade do Minho, 4710-057 Braga, Portugal}
\affiliation{Departamento de
  Astronom\'{\i}a y Astrof\'{\i}sica, Universitat de Val\`encia,
  Dr. Moliner 50, 46100, Burjassot (Val\`encia), Spain}

\author{Juan Calder\'on~Bustillo}
\email{juan.calderon.bustillo@gmail.com}
\affiliation{Instituto Galego de F\'{i}sica de Altas Enerx\'{i}as, Universidade de
Santiago de Compostela, 15782 Santiago de Compostela, Galicia, Spain}

\author{Jos\'e~A. Font}
\email{j.antonio.font@uv.es}
\affiliation{Departamento de
  Astronom\'{\i}a y Astrof\'{\i}sica, Universitat de Val\`encia,
  Dr. Moliner 50, 46100, Burjassot (Val\`encia), Spain}
\affiliation{Observatori Astron\`omic, Universitat de Val\`encia,  Catedr\'atico 
  Jos\'e Beltr\'an 2, 46980, Paterna (Val\`encia), Spain}

\author{Solange~Nunes}
\email{solangesilnunes@gmail.com}
\affiliation{Centro de F\'{\i}sica das Universidades do Minho e do Porto (CF-UM-UP), Universidade do Minho, 4710-057 Braga, Portugal}

\author{Antonio~Onofre}
\email{antonio.onofre@cern.ch}
\affiliation{Centro de F\'{\i}sica das Universidades do Minho e do Porto (CF-UM-UP), Universidade do Minho, 4710-057 Braga, Portugal}

\author{Alejandro \surname{Torres-Forn\'e}}
\email{alejandro.torres@uv.es}
\affiliation{Departamento de
  Astronom\'{\i}a y Astrof\'{\i}sica, Universitat de Val\`encia,
  Dr. Moliner 50, 46100, Burjassot (Val\`encia), Spain}
\affiliation{Observatori Astron\`omic, Universitat de Val\`encia,  Catedr\'atico 
  Jos\'e Beltr\'an 2, 46980, Paterna (Val\`encia), Spain}

\begin{abstract}
We evaluate several neural-network architectures, both convolutional and recurrent, for gravitational-wave time-series feature extraction by performing point parameter estimation on noisy  waveforms from binary-black-hole mergers. We build datasets of 100,000 elements for each of four different waveform models (or approximants) in order to test how approximant choice affects feature extraction. Our choices include  \texttt{SEOBNRv4P} and \texttt{IMRPhenomPv3}, which contain only the dominant quadrupole emission mode, alongside \texttt{IMRPhenomPv3HM} and \texttt{NRHybSur3dq8}, which also account for high-order modes. Each dataset element is injected into detector noise corresponding to the third observing run of the LIGO-Virgo-KAGRA (LVK) collaboration. We identify the Temporal Convolutional Network (TCN) architecture as the overall best performer in terms of training and validation losses and absence of overfitting to data. Comparison of results between datasets shows that the choice of waveform approximant for the creation of a dataset conditions the feature extraction ability of a trained network. Hence, care should be taken when building a dataset for the training of neural networks, as certain approximants may result in better network convergence of evaluation metrics. However, this performance does not necessarily translate to data which is more faithful to numerical relativity simulations. We also apply this network on actual signals from LVK runs, finding that its feature-extracting performance can be effective on real data.
\end{abstract}

\maketitle

\section{Introduction}
The third observing run of the LIGO-Virgo-KAGRA (LVK) collaboration delivered a total of 79 gravitational-wave (GW) events, raising the total number of detected signals to  date to 90~\cite{abbott_gwtc-1_2019,abbott_gwtc-2_2020,LIGOScientific:2021usb,the_ligo_scientific_collaboration_gwtc-3_2021}. All these events are consistent with  compact binary coalescences (CBCs) comprising mostly black-hole mergers (BBH) but also neutron-star mergers \cite{TheLIGOScientific:2017qsa,GW190425} and neutron-star black-hole mergers \cite{NSBHs}. 
The detection of signals from CBCs as well as the extraction of physical information of the sources relies on matched filtering \cite{MatchedFilter}. This is based on the comparison (through cross-correlation) of the incoming GWs to pre-computed signal templates (or waveforms) for given source parameters \cite{FindChirp,Usman:2015kfa,Veitch:2014wba,Ashton:2018jfp}. This process requires that templates are faithful representations of true GWs, as the opposite can introduce biases in the source parameter estimation or even actual misses of incoming signals.

The field of waveform modelling has gradually reached a status of maturity, currently providing robust methods for the description of gravitational radiation from compact binary systems through the combination of information from three main physical frameworks, namely post-Newtonian theory, the effective-one-body approach, and numerical relativity. There exist nowadays accurate and efficient waveform models, or {\it approximants}, describing all stages of the signal of a BBH coalescence (inspiral, merger and ringdown) and allowing for accurate parameter estimation of the source. This is especially true for quasi-circular binaries, also accounting for the effects of unequal masses, misaligned spins, precession, and subdominant (high-order) modes \cite{khan_frequency-domain_2016,khan_phenomenological_2019,khan_including_2020, varma_surrogate_2019, bohe_improved_2017,cao_waveform_2017,akcay_hybrid_2020, nagar_time-domain_2018}. 

Given a waveform model, the traditional method to infer source parameters is through Bayesian inference. This is a computationally expensive task as it requires  applying  stochastic  sampling  techniques to evaluate the likelihood function for a large set of source parameters, resulting in the need to generate potentially millions of waveforms.
Moreover, as the performance of GW detectors increases, with predicted sensitivities for third-generation detectors being around a tenfold improvement on the current instruments \cite{belgacem_cosmology_2019}, the amount of detected signals is expected to increase accordingly. This may render unsustainable traditional  approaches for both, signal detection and parameter estimation. Recently, there have been successful attempts to significantly speed up inference using machine-learning techniques ~\cite{Green:2020,dax_real-time_2021,williams_nested_2021,bayley_rapid_2022,Gabbard:2022,Dax:2023,bhardwaj2023peregrine}.
The role of deep learning (DL) for GW data analysis is already relevant and is bound to become increasingly so, as GW astronomy fully unfolds (see~\cite{Huerta:2019,Cuoco:2020} and references therein). Current applications are not only limited to accelerating parameter estimation but also include efforts to develop detection methods \cite{alvares_exploring_2021, ALBUS, mockdatachal}, to improve signal quality \cite{torres-forne_denoising_2016,torres-forne_application_2020}, to accelerate waveform generation \cite{stefano_mlgw, stefano_bns} as well as to simulate noise transients in GW detectors \cite{melissa_gans,lopez_gengli}. 

There are several relevant choices when implementing a DL method to deal with GW data. On the one hand, there is the choice of a network architecture. There is already a rich body of work when it comes to the application of DL to generic time series \cite{fawaz_deep_2018,lara-benitez_experimental_2021,tan_time_2021}. Therefore, there is a large amount of pre-existing architectures that can detect relevant features of time series data (or at least are able to serve as the basis for new architectures). On the other hand, however, there is the issue of having an appropriate dataset. Points to consider include data representation, such as the window size in time and the sample rate of the data, as well as more involved technical aspects such as dimensional reduction through, for example, principal component analysis. More fundamentally, though, there is the issue of the composition of the dataset. The number of confirmed GW detections is minuscule by data science standards to realistically attempt to train models on real data. Moreover, full numerical relativity (NR) simulations are too computationally expensive to cover the entire parameter space of BBH coalescences. As such, waveforms for use in GW astronomy are commonly generated by approximants which attempt to quickly and accurately describe the waveform generated by the full NR simulation, developed with different physical approximations in mind, and then calibrated to NR waveforms \cite{santamaria_matching_2010,khan_frequency-domain_2016,nagar_time-domain_2018,cao_waveform_2017,bohe_improved_2017,akcay_hybrid_2020, cotesta_frequency_2020}. 

The existence of different approximants relying on different techniques raises the question of whether choosing a particular one for the creation of a dataset may condition the feature extraction ability of a trained network.
The aim of this paper is to address this question. To do so, we test a number of state-of-the-art network architectures implemented in the \texttt{tsai} python package \cite{oguiza_tsai_2022} on the task of point parameter estimation, as a way to evaluate how well the waveform's relevant features are identified. Our goal is to find out if it is possible to select a specific type of architecture that might  extract the features of the data relevant to GWs in a more accurate way than the rest. Furthermore, after finding the best-performing architecture, we test the effect of approximant choice on the network performance. In particular, we analyse the differences between a NR surrogate model, \texttt{NRHybSur3dq8}, the phenomenological model \texttt{IMRPhenomPv3} and its sibling \texttt{IMRPhenomPv3HM} including high-order modes, and the effective-one-body based model \texttt{SEOBNRv4P}. Our results indicate that the choice of approximant to build GW datasets can have a non negligible impact on feature extraction.

The paper is organized as follows: In Section II we describe the generation of the four datasets we use for this work. In Section III we compare the training of 10 different networks on the \texttt{NRHybSur3dq8} dataset, choosing the network with the lowest validation loss as the best architecture for our purposes. In Section IV we take this architecture and train 10 equal networks on each of the generated datasets and compare the results by looking at the root-mean-square error (RMSE) of each of the physical parameters studied. Finally, in Section V we summarize  our conclusions.

\section{Datasets}

\subsection{Approximant overview}

The first approximant we employ in our study is \texttt{IMRPhenomPv3}~\cite{khan_phenomenological_2019}. This is a phenomenological waveform approximant in the \texttt{IMRPhenomP} family~\cite{Schmidt:2015}. This family of approximants is constructed by matching NR simulations of BBH systems to a set of physically-motivated analytical waveforms, typically in the frequency domain, including terms that attempt to describe the full physics of the dominant $(\ell,m)=(2,2)$ mode of compact binary gravitational radiation, encompassing spin effects such as precession. These approximants are then calibrated to a set of NR simulations, in order to obtain a good agreement with the predictions of general relativity. We also employ a second approximant of the same family, \texttt{IMRPhenomPv3HM}~\cite{khan_including_2020} which builds upon the foundations of \texttt{IMRPhenomPv3} but includes higher (subdominant) mode effects, specifically the (2,2), (2,1), (3,3), (3,2), (4,4), and (4,3) modes. The inclusion of these effects was shown to significantly improve the mismatch to NR waveforms for systems with mass ratios up to $1/5$.

The third approximant we use is \texttt{SEOBNRv4P}~\cite{bohe_improved_2017}. This is part of a family of approximants which utilise the effective-one-body (EOB) formalism. EOB describes the two-body dynamics in general relativity and its GW emission using a resummation of the post-Newtonian information for the dynamics in terms of the geodesic motion of a particle in an effective spacetime~\cite{Buonanno:1999}. The two-body problem is mapped into the effective problem of a particle of mass $\mu$ (the reduced mass of the binary system) in an effective spacetime which is a symmetric deformation of Schwarzschild (or Kerr for spinning binaries). The conservative dynamics of the two-body system is encoded in the EOB Hamiltonian from which the evolution of the phase space variables can be obtained. First attempts to construct a resummed version of the GW flux incorporating test-particle results at higher order were taken by~\cite{Damour:1998}. A different strategy based on a resummation framework for the GW amplitudes was subsequently initiated by~\cite{Damour:2008}. Recent comparisons with NR simulations and test-particle results outperform the standard Taylor-expanded form approximants~\cite{Nagar:2019}. 

Our fourth and last approximant is \texttt{NRHybSurd3q8}~\cite{varma_surrogate_2019}. This is a hybrid surrogate of NR waveforms. Surrogate models work by interpolating the space of gravitational waveforms using a reduced basis obtained from a number of NR waveforms. However, despite their high accuracy, surrogate models have the disadvantage of NR waveforms being rather short, usually starting around twenty orbits before the BBH merger. \texttt{NRHybSur3dq8}  gets around this issue by stitching together EOB-corrected post-Newtonian waveforms for the early inspiral with corresponding NR waveforms, and then applying the surrogate methodology to obtain a model that is capable of accurately interpolating the space of waveforms while still being able to reproduce a large portion of the inspiral phase.

\subsection{Dataset Generation}

\begin{figure*}[t]
\centering
\includegraphics[width=0.8\textwidth]{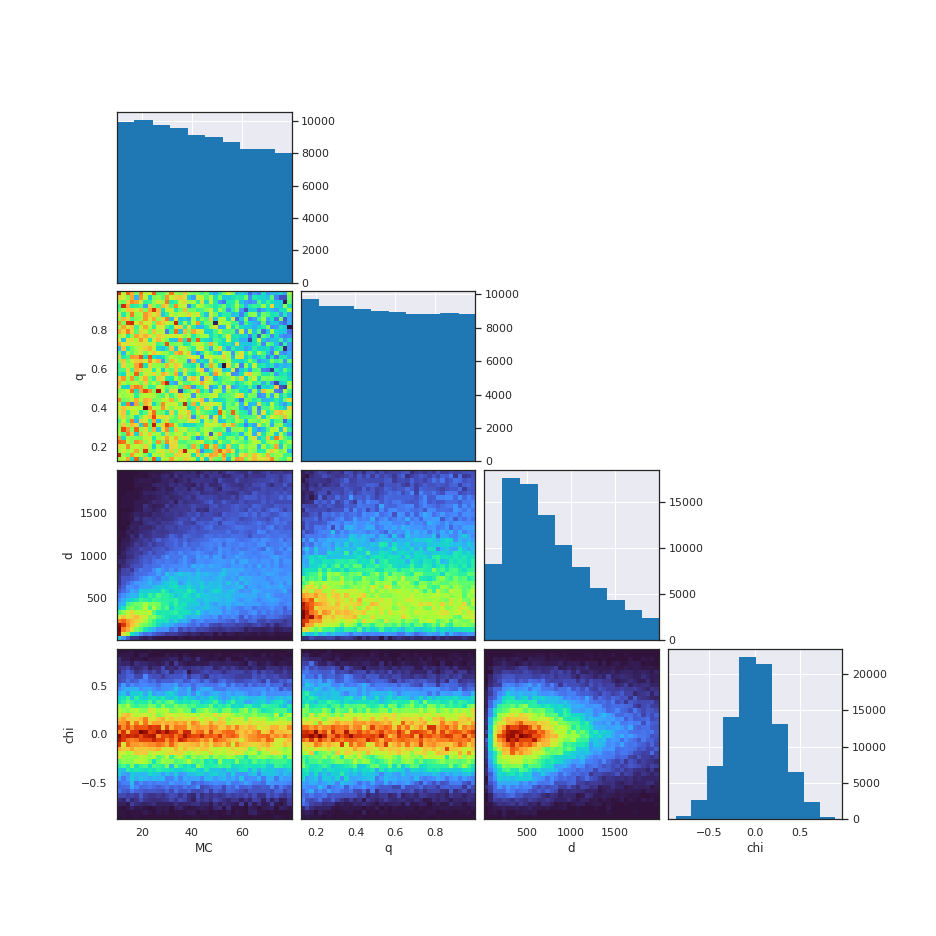}
\caption{\label{fig:datasets} Dataset composition for the 100,000 injections using the \texttt{NRHybSur3dq8} approximant. The composition of the datasets using other approximants is not shown as it is equivalent to the one displayed in the figure. 
}
\end{figure*}

The datasets are generated in two steps. First, a dataset of GW waveforms corresponding to 100,000 BBH mergers is created for each of the four chosen approximants. This is done using the \texttt{pycbc} package \cite{usman_pycbc_2016} to generate the signals given a dictionary of seven physical parameters. These parameters are sampled as follows:
the primary detector-frame mass of the system, $m_1$, is sampled uniformly in the interval $[5, 80]M_\odot$. 
The mass ratio, $q$, is sampled uniformly in the interval $[1/8, 1]$, in order to respect the limitations of the \texttt{NRHybSur3dq8} model. 
The secondary mass, $m_2$, is taken to be the maximum value between $5M_\odot$ and $(q\cdot m_1) M_\odot$.
The spins of the component objects, $s_1$ and $s_2$, have their dimensionless magnitudes sampled uniformly in the interval $[0,0.9]$, while their direction is sampled isotropically. However, the $x$ and $y$ components are  ignored in order to respect the parameter space of the hybrid surrogate model.
The inclination $\iota$ is sampled uniformly in the range $[0, \pi]$.  
Finally, the coalescence phase is sampled uniformly from $[0, 2\pi]$.
All of these waveforms are generated with an initial nominal distance of 100 Mpc, which serves the purpose of scaling them. This distance is then adjusted  during the injection process according to the selected signal-to-noise (SNR) criterion (see below). 

The second step comprises the injection of the generated waveforms into real detector strain for each of the LIGO detectors \cite{aligo2015} and Virgo \cite{avirgo2014}, taken from a 6.5-hour time interval from the O3 observation run, starting at GPS time 1264316210. No GW signals have been reported by the LVK Collaboration in this interval, which ensures only real noise conditions. In this study we do not worry about the effects possible glitches in the data might cause to our analysis. This means that, while real parameter estimation analysis are carried our after checking that no glitches are present in the data, our analysis is subject to biases due to potential superposition of glitches over injections, therefore providing conservative results. The injection is performed using the \texttt{bilby} package for python~\cite{bilby}. An important choice made at this stage was to limit the detector network's SNR (given by the quadrature sum of the individual detector SNRs) to the interval $[10, 50]$. This guarantees some minimal signal quality without going into unrealistically high values. To this end, for each approximant used, the process of injection is as follows:
\begin{enumerate}
\item A waveform from a generated dataset is loaded.
\item A target network SNR is sampled uniformly in the range $[10, 50]$. 
\item A 20 second segment of real strain data for each of the detectors (LIGO Hanford, LIGO Livingston \cite{aligo2015} and Virgo \cite{avirgo2014}) is loaded, with a randomly chosen GPS time serving as the initial time.
\item The sky position of the source is sampled isotropically.
\item The waveform is injected into the detector strains at $t=t_0+10$, with an additional random value up to $0.8$ 
\item The injection SNR is calculated as 
\begin{eqnarray}
\text{SNR}=\frac{(h|s)}{\sqrt{(h|h)}}\,,
\end{eqnarray}
 where $h$ stands for the template waveform and $s$ stands for the post-injection strain data. Note that here we use the conventional noise-weighted inner product definition, given a frequency-series $g(f)$ and $h(f)$
\begin{eqnarray}
(g|h) = 2\int_0^\infty\frac{g^*(f)h(f)+g(f)h^*(f)}{S_n(f)}df\,, 
\end{eqnarray}
with  $S_n$ being the one-sided power spectral density of the detector noise.
\item The injection SNR is compared to the target SNR. If the absolute difference is less than 3 the final waveform is accepted and written to the injection dataset. If this criteria is not met, a ratio 
\begin{eqnarray}
r=\frac{\mathrm{target \,SNR}}{\mathrm{SNR}}
\end{eqnarray}
is calculated and step 6 is repeated with the luminosity distance of the clean waveform being scaled by $1/r$. 

\item Steps 1-7 are repeated until all waveforms from a given dataset have been successfully injected.
\end{enumerate}
The resulting injections (at a sample rate of $4096~$Hz) and all the physical parameters used to generate the original waveforms are saved to an HDF5 file. To perform our study, however, we will focus solely on 4 parameters:  the (detector-frame) chirp mass of the system 
\begin{eqnarray}
\mathcal{M}=\frac{(m_1 m_2)^{3/5}}{(m_1+m_2)^{1/5}}\,,
\end{eqnarray}
the mass ratio $q={m_2}/{m_1}$, the effective inspiral spin 
\begin{eqnarray}
\chi_{\textrm{eff}}=\frac{m_1\cdot {s^z}_1+m_2 \cdot {s^z}_2}{m_1 + m_2},
\end{eqnarray}
where $s^1_z$ and $s^2_z$ are the components of the objects' spins aligned with the orbital angular momentum, and the luminosity distance $d_L$. We deliberately do not cover the entire parameter space. These four parameters are chosen due to their larger influence on the gravitational waveforms and, as such, differences in feature extraction performance should be more obvious in this reduced parameter space. 

The distribution of these parameters for the \texttt{NRHybSur3dq8} dataset is displayed in \cref{fig:datasets}. {Note that the imposition of a flat SNR distribution results, effectively, in a prior in distance proportional to $1/d_L$. This is evident in \cref{fig:datasets} and it does mean that the network will have a bias along those lines. On the other hand, typical priors on the distance go with $d^2_L$. We justify this by again pointing out that our goal is not application-level parameter estimation but rather to test feature extraction, and setting the SNR as described makes network convergence more robust}.

\section{Network architecture overview}

For this work we implement the neural networks using the \texttt{fastai} package (version \texttt{2.7.10}) for python~\cite{howard_fastai_2018} (version \texttt{3.8.13}), which functions as a  while the architectures are taken from the time-series-focused \texttt{tsai} package~\cite{oguiza_tsai_2022} (version \texttt{0.3.5}). In total we test 10 different networks, 4 of which are convolutional networks and 3 of which are recurrent networks. Most of the networks we consider were designed for classification tasks, save for the recurrent networks which see the most use in forecasting. This choice of networks architectures was pragmatic: given the rapid pace and subject breadth of DL research, finding a set of networks which stands out in a general way is probably unfeasible. As such, we chose to include recurrent architectures as they are a mainstay in time-series forecasting, as well as some of the top-performing convolutional networks in \cite{tan_time_2021}. 

\subsection{Convolutional networks}

The sample of convolutional networks we employ is the following:

    \begin{itemize}
        \item Fully Convolutional Network (FCN)~\cite{wang_time_2016}: the FCN architecture makes use of simple convolution operations in sequence in order to extract the features of the data, followed by a linear layer which maps the convolutions' output to the desired output size (in our case, the number of parameters).

        \item xResNet1d: The xResNet1d architecture is a tweaked version of the ResNet architecture \cite{he_deep_2015}. ResNets are convolutional-based networks which make use of residual connections between layers, that is, the original input of the layer is added to the output of a convolution operation happening within the layer. To be more precise, for a layer input $\mathbf{x}$, the layer output is $f(\mathbf{x})=\mathbf{x}+\mathcal{C}(\mathbf{x})$, where $\mathcal{C}$ stands for the convolutional operation within the layer. The xResNet architecture tweaks the original ResNets, avoiding some information loss by delaying feature reduction operations and replacing expensive convolutions with multiple cheap convolutions~\cite{he_bag_2018}. We use three different depths for this network, namely 18, 34, and 50 layers, due to our previous positive results for GW parameter estimation using xResNets obtained in \cite{alvares_exploring_2021}.
   
        \item InceptionTime~\cite{fawaz_inceptiontime_2020}: The InceptionTime architecture is at heart a residual network. Its main distinguishing feature is the use of the Inception block, which combines concurrent information on the various channels of a multivariate time series in order to obtain a univariate representation of the original data.
        
        \item Temporal Convolutional Network (TCN)~\cite{bai_empirical_2018}: the TCN architecture is once again typically based on residual blocks. However, TCN stands out by the use of causal convolutions, that is, throughout the network elements of a layer are considered to be in sequence and are only affected by the elements of the previous layer which are in the past. TCN also employs dilated convolutions, so that deeper layers can have a larger receptive field, i.e. they can look further into the past.
        
        \item MiniRocket~\cite{dempster_minirocket_2021}: MiniRocket is a convolutional network for time series which  performs convolutions using a set of 84 convolutional kernels whose elements have their values restricted to the set $\{1,2\}$. Using randomly generated bias values, MiniRocket generates 9,960 features per timeseries, which are then passed to a linear layer to obtain the final output.
    \end{itemize}  
    
\subsection{Recurrent networks}    
    
Correspondingly, our sample of recurrent networks comprises the following ones:
    
    \begin{itemize}
        \item Recurrent Neural Network (RNN)~\cite{rumelhart_learning_1987}: RNN makes use of memory cells which keep track of the network's immediate past state during training, using it as one of the parameters that defines the next step in the network's parameter space.
        
        \item Long Short-Term Memory (LSTM)~\cite{hochreiter_long_1997}: LSTMs are an extension of the RNN framework. However, they address a weakness in the simple RNN setup. RNNs have difficulty incorporating long-term dependencies, as gradients of older states often either vanish (the most common case) or explode as they propagate. Under the LSTM setup, the network is endowed with ``input", ``output" and ``forget" gates that allow it to control information flow, for example being able to ignore potential updates to its memory cells if the input is deemed irrelevant, and being able to forget contents of its memory. As such it is less susceptible (though not immune) to the vanishing/exploding gradients problem.
        
        \item Gated Recurrent Unit (GRU)~\cite{chung_empirical_2014}: the GRU architecture adopts the information flow modulation idea from the LSTM concept, but crucially GRU networks do not have an equivalent to the output gate, which modulates the contents of the memory cell that get exposed to the other components of the network. This makes the network computationally simpler but potentially less able to detect long-distance correlations.
        
    \end{itemize}
    
\begin{figure}
    \centering
    \includegraphics[width=\linewidth]{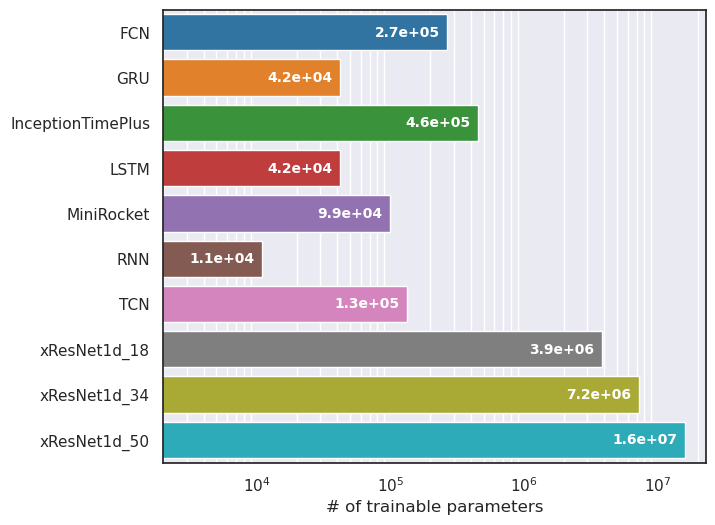}
    \caption{Number of trainable parameters for the different networks tested.}
    \label{fig:nmp_par}
\end{figure}  

\begin{figure*}
    \begin{subfigure}[b]{0.495\textwidth}
    \centering
    \includegraphics[width=\linewidth]{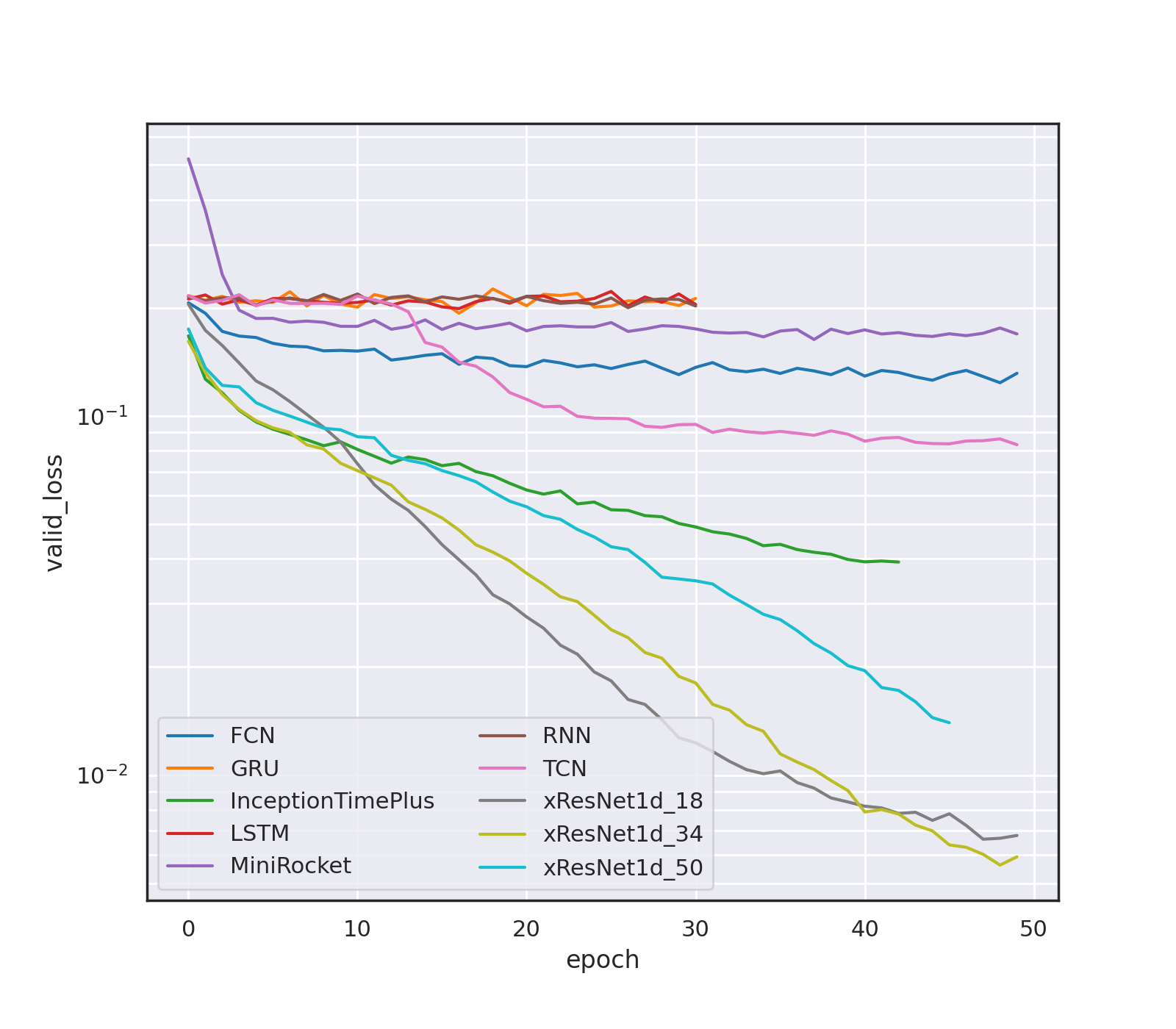}
    \caption{ \label{fig:tlosses}}
    \end{subfigure}
    \begin{subfigure}[b]{0.495\linewidth}
    \centering
    \includegraphics[width=\linewidth]{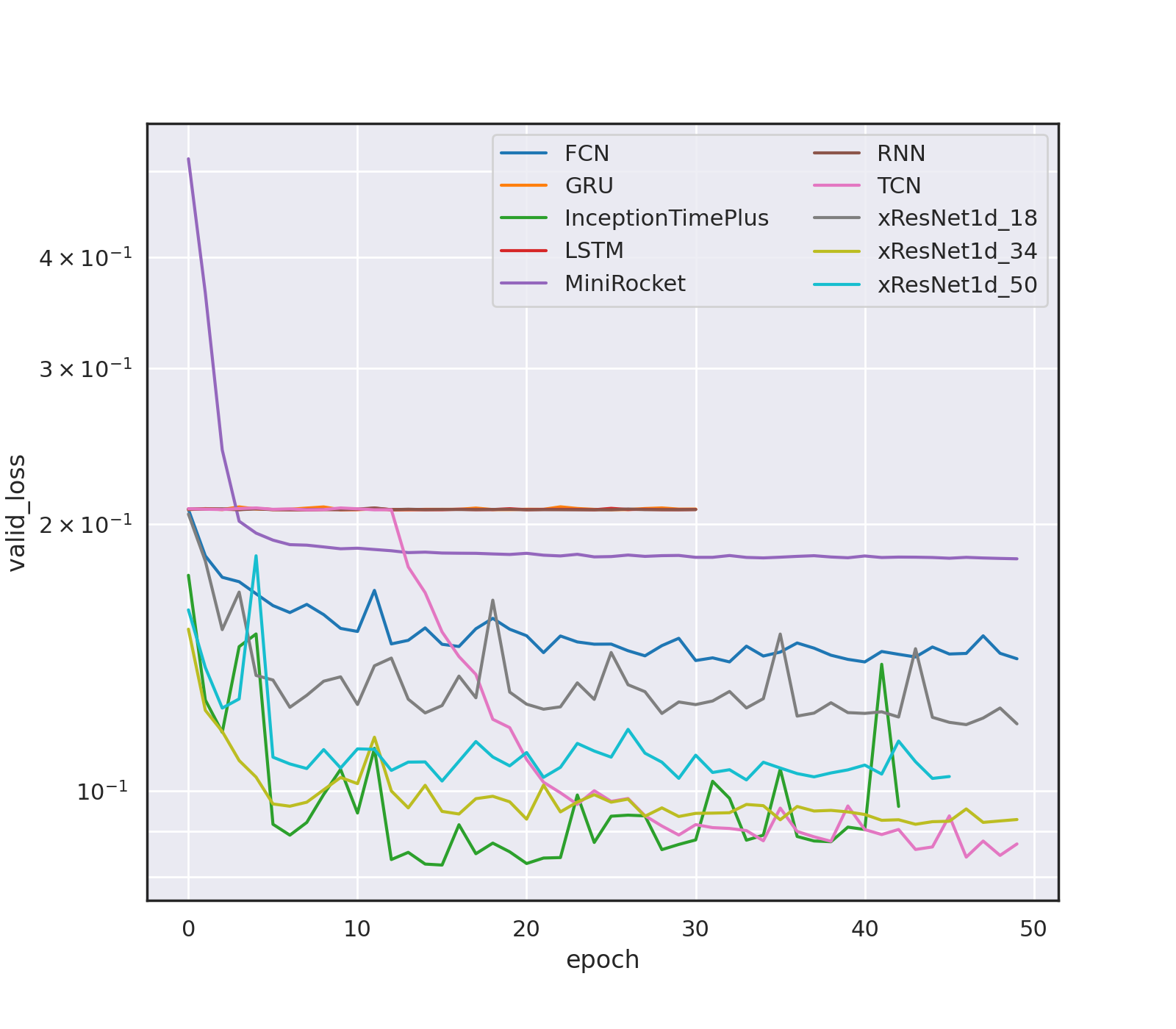}
    \caption{ \label{fig:vlosses}}
    \end{subfigure}
    \caption{Evolution of the loss function during the training of the networks using the GW dataset built with the \texttt{NRHybSur3dq8} approximant. \textbf{(a):} training loss. \textbf{(b):} validation loss. \label{fig:losses}}
\end{figure*}

\section{Results}

For all of the networks we use the default architecture parameters provided by \texttt{tsai}, changing only the number of input and output channels to match the data and the number of parameters estimated, respectively. The different networks have a rather large number of trainable parameters and, moreover, their amounts vary noticeably among networks.  This is displayed in \cref{fig:nmp_par} which shows that the number of parameters to train can be as low as $1.1\times 10^4$ for the case of {RNN} and as high as $1.6\times 10^7$ for {xResNet1d} with 50 layers.

\subsection{Training and architecture comparison}

We train each of our ten networks on the dataset built with the \texttt{NRHybSur3dq8} approximant\footnote{All computations in this paper were performed on the Artemisa cluster from the University of Valencia using 32 Intel Xeon CPU cores and an Nvidia V100 GPU.} which belongs to the class of waveform models that yield the most accurate results for quasi-circular BBH systems with aligned spins. We
only show the comparison of the training of the different network architectures for this approximant  as a similar behaviour  is found for the datasets built with the other three approximants regarding the evolution of the loss function. 

We prepare the data before passing it to the network, rescaling the ${\cal M}$ and $d_L$ values so that they fit in the interval $[0,1]$. To evaluate the parameter values predicted by the network in comparison to the real injection parameters in the dataset we use a simple mean-square-error (MSE) loss function, given by 
\begin{eqnarray}
\mathrm{MSE}(x_\mathrm{in},y_\mathrm{truth},\Theta) = \frac{\sum_i [y_\mathrm{truth}^i-f(x_\mathrm{in}^i, \Theta)]^2}{N},\,
\end{eqnarray} where $N$ is the number of inputs passed to the network, $y_\mathrm{truth}$ is a 2D $(N_\mathrm{params} \times N)$ tensor of ground truth values, $x_\mathrm{in}$ is a 3D array of input tensors with dimensions $([L_\mathrm{signal} \times N_\mathrm{channels}] \times N)$, where $L_\mathrm{signal}$ is the length of each signal, $N_\mathrm{channels}$ is the number of channels in the multivariate time series (that is, in this case, the number of GW detectors considered), $f$ is the mapping represented by the network, taking each of the $N$ $[L_\mathrm{signal} \times N_\mathrm{channels}]$-shaped input tensors to a 1D tensor of length $N_\mathrm{params}$, and $\Theta=(\theta_0, \theta_1, ...)$  is the array of learned weights which define $f$.
We also use $L_2$ regularization, which means the square of the weights of the network are added to the loss function with some chosen multiplier ($10^{-5}$, in our case), making the full loss function take the form
\begin{eqnarray}
\mathcal{L}(x_\mathrm{in},y_\mathrm{truth},\Theta) = \mathrm{MSE}(x_\mathrm{in},y_\mathrm{truth},\Theta)+\frac{10^{-5}}{2}\Theta^2\,.
\end{eqnarray}
This regularization method prevents the network from giving too much importance to a limited set of weights.
For the training process, we make use of the well-established Adam optimizer \cite{kingma_adam_2014}.  Using a batch size of 256 and a 0.8/0.2 training/validation split, we start the training process by using a learning rate finder, performing a sweep on learning rate values, starting at $10^{-6}$ and monotonically increasing with each batch evaluated, up to the point where the learning rate is high enough that the loss function value diverges. We obtain a starting learning rate by taking the learning rate value corresponding to the lowest MSE loss in this process. 
We train for a maximum of 100 epochs, using an early stopping call if the validation loss does not decrease by at least 0.002 after 20 epochs. We also reduce the learning rate by a factor of 10 after 5 epochs if there is not a validation loss decrease of at least 0.005 in that period. 
\begin{figure*}
\begin{subfigure}[b]{0.495\textwidth}
        \centering
        \includegraphics[width=\linewidth]{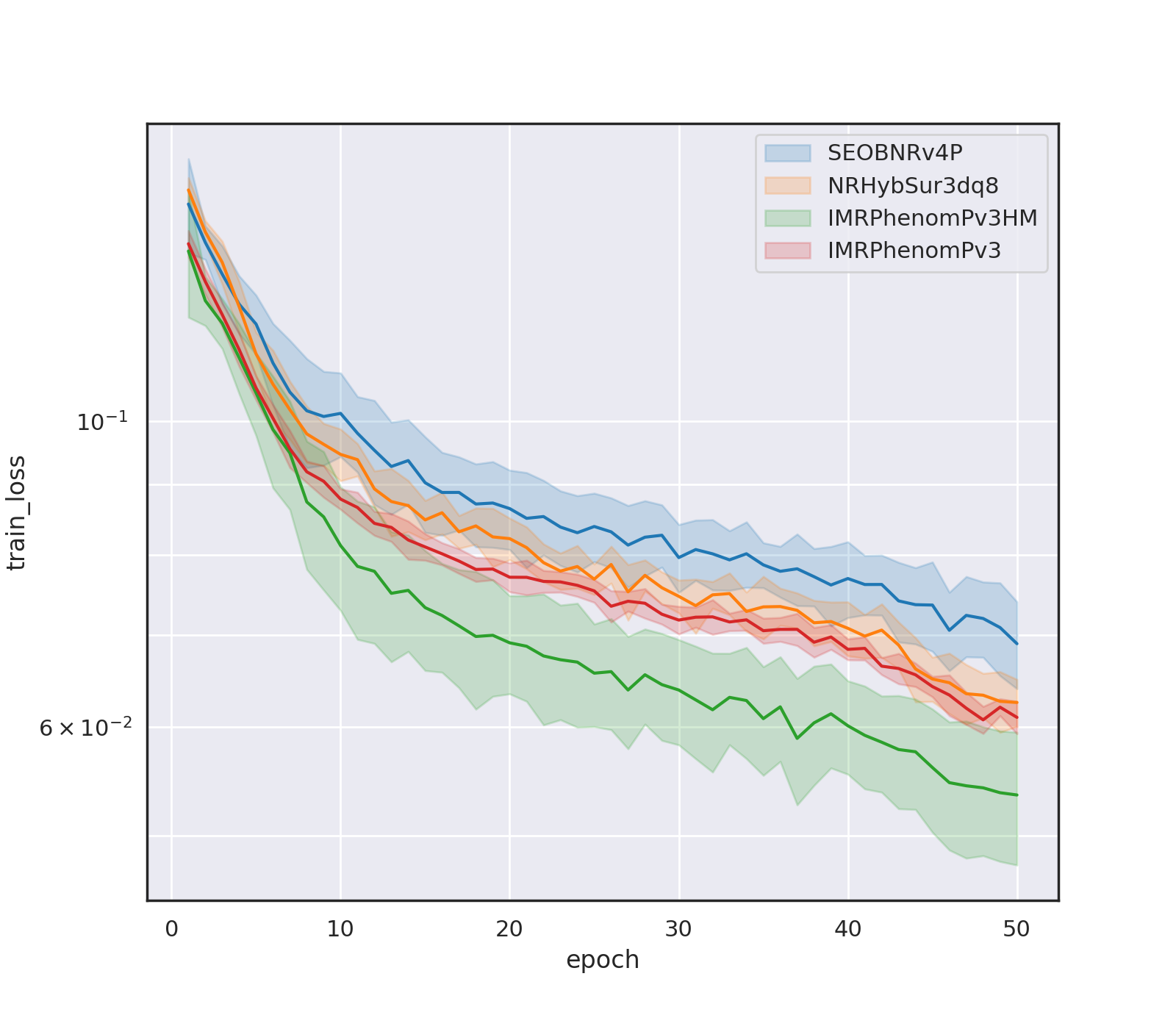}
        \caption{\label{fig:avg_tloss}}
    \end{subfigure}
    \begin{subfigure}[b]{0.495\textwidth}
        \centering
        \includegraphics[width=\linewidth]{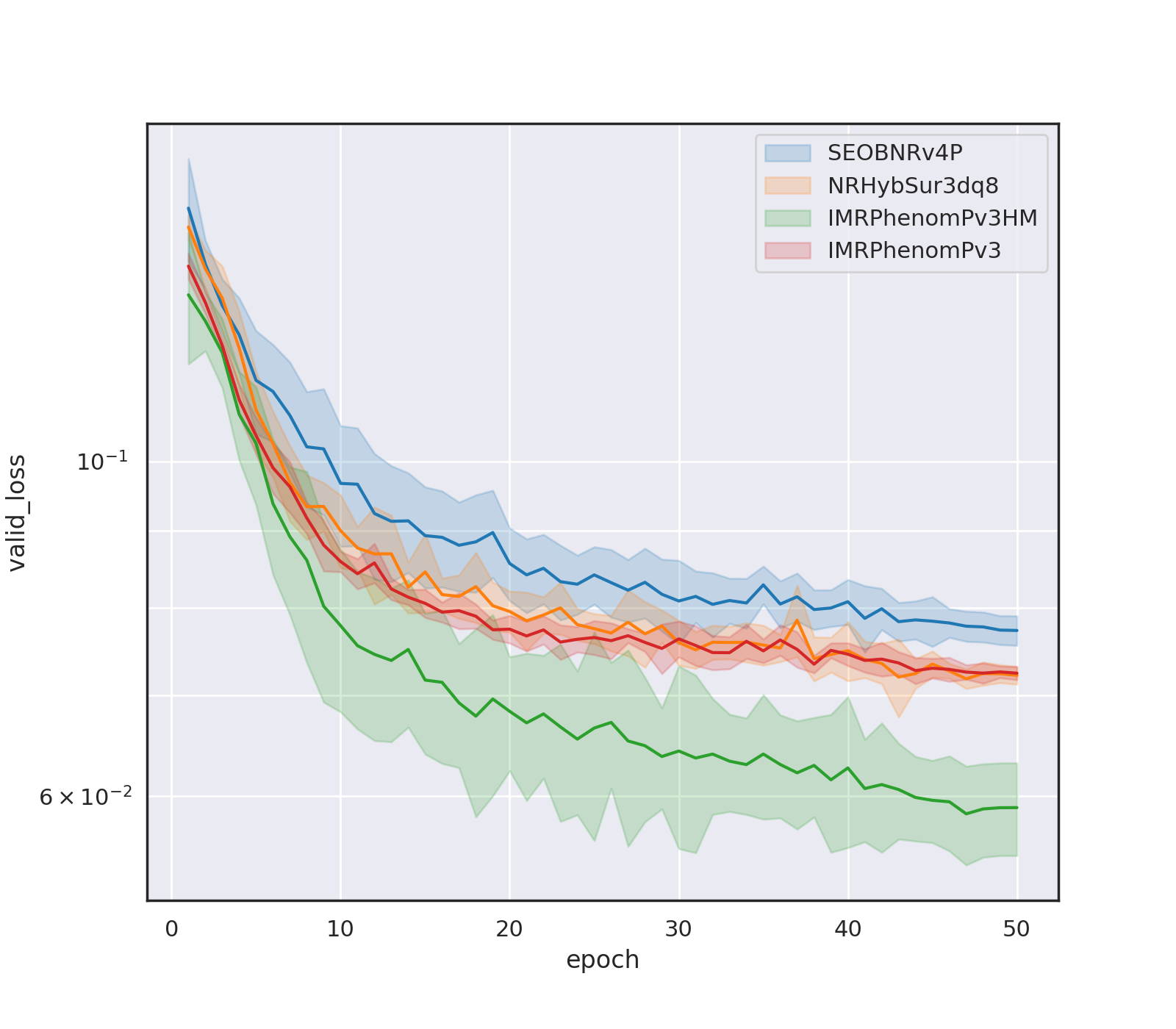}
        \caption{\label{fig:avg_vloss}}
    \end{subfigure}
    \caption{Evolution of the loss function during training of the networks for the approximant comparison test. Lines represents the mean for all the networks trained on a given approximant, while the shaded area represents the standard deviation interval at each epoch. \textbf{(a):} training loss. \textbf{(b):} validation loss. \label{fig:avg_train}}
\end{figure*}

The evolution of the two losses during training for each network are presented in \cref{fig:losses}. The most relevant metric here is the validation loss (right panel of the figure) as it approximately represents the network's performance on out-of-distribution data. Interestingly, the recurrent architectures struggle to converge at all for the RNN and LSTM cases (for the RNN this may indicate gradient vanishing or explosion, as discussed in \cite{DBLP:journals/corr/abs-1211-5063}, but for the LSTM case such effects should be mitigated), yet the GRU architecture presents the second-lowest validation loss overall, showing no overfitting, despite not having a particularly significant difference in the number of parameters. Further attempts at making networks based on the RNN and LSTM architectures converge, by fine-tuning the learning rate and some network hyperparameters, yielded no improvement. This issue may require further study. As for the convolutional architectures, while the InceptionTime and xResNet architectures show promising behaviour at the start of training, they quickly enter into an overfitting regime, with the validation losses showing unstable behaviour and values of an order of magnitude higher than the corresponding training losses. The network based on the MiniRocket architecture shows a consistently decreasing validation loss, but this process is rather slow and ends up being noncompetitive given the restriction on the number of epochs. The FCN-based network does not seem to converge, rapidly showing unstable behaviour in the training loss. The TCN-based network maintains a consistent improvement until starting to stabilise at around epoch forty, and the closeness between training and validation loss values indicates that no significant overfitting or underfitting is occurring. From the final value of the validation loss (aproximately $2\times10^{-2}$) and the lack of overfitting or underfitting, we thus conclude that the best performing network of our sample is the one using the TCN architecture.

\begin{figure*}

    \centering
    \begin{subfigure}[b]{0.45\textwidth}
        \includegraphics[width=\textwidth]{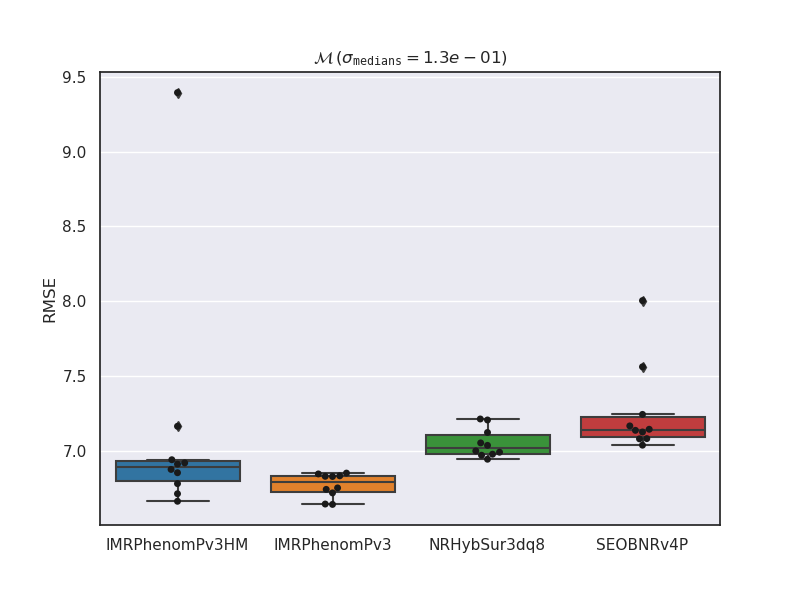}
        \caption{}
    \end{subfigure}
    \begin{subfigure}[b]{0.45\textwidth}
        \includegraphics[width=\textwidth]{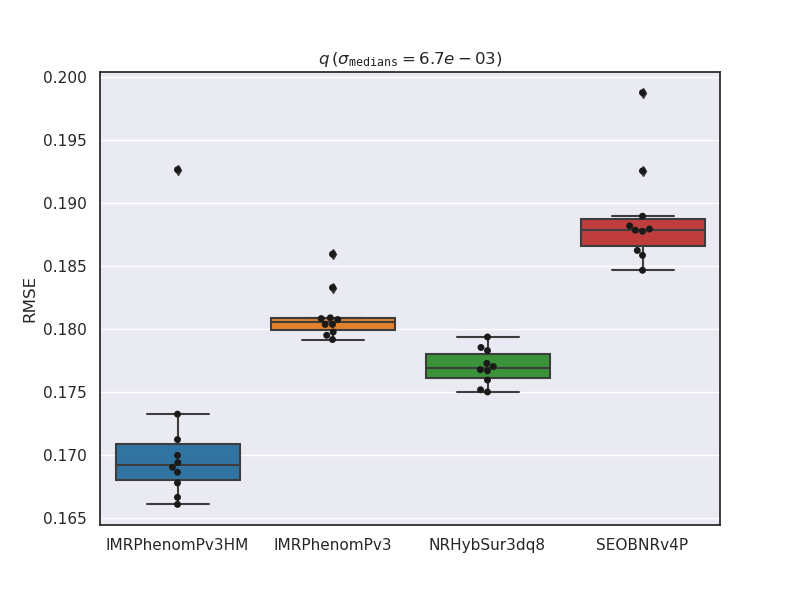}
        \caption{}
    \end{subfigure}
    \begin{subfigure}[b]{0.45\textwidth}
        \includegraphics[width=\textwidth]{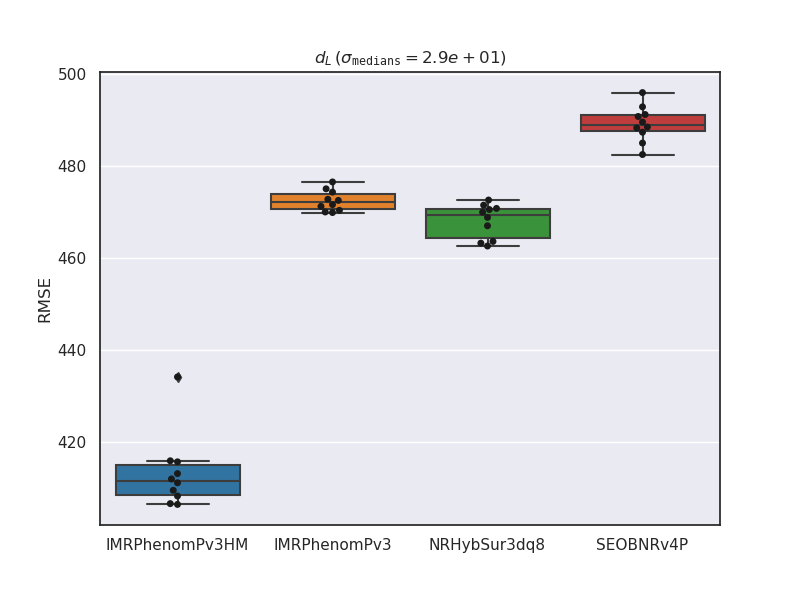}
        \caption{}
    \end{subfigure}
    \begin{subfigure}[b]{0.45\textwidth}
        \includegraphics[width=\textwidth]{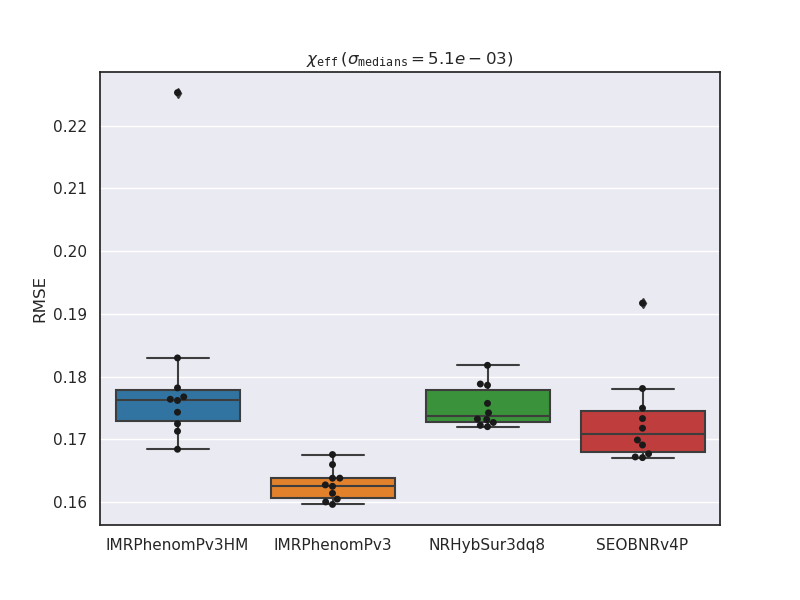}
        \caption{}
    \end{subfigure}
    \caption{Box plots for the RMSE of the TCN networks on the validation set, for each of the physical parameters used to train the network and for each of the four datasets used. The specific panels show the RMSE for \textbf{(a)} the chirp mass of the system, \textbf{(b)} the mass ratio,  \textbf{(c)} the luminosity distance, and \textbf{(d)} the  effective inspiral spin of the system.} 
    \label{fig:boxplots}
\end{figure*}

\subsection{Approximant comparison}
\begin{figure*}

    \begin{subfigure}[b]{0.495\textwidth}
        \centering
        \includegraphics[width=\linewidth]{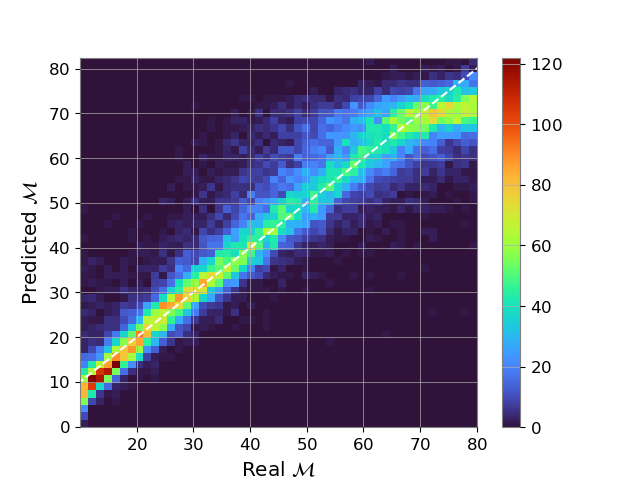}
        \caption{\label{fig:IMR_M}}
    \end{subfigure}
    \begin{subfigure}[b]{0.495\textwidth}
        \centering
        \includegraphics[width=\linewidth]{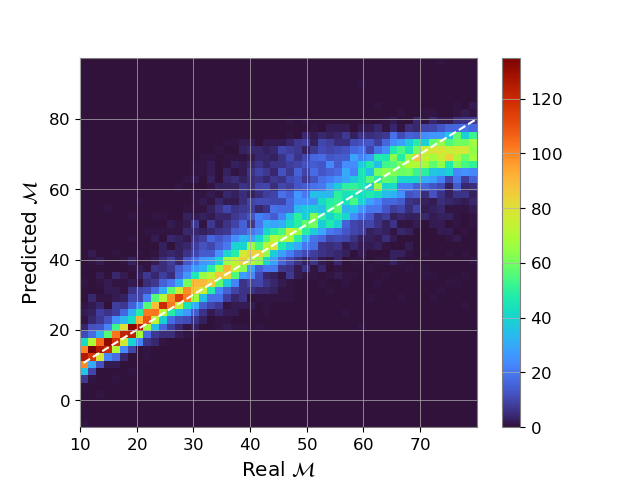}
        \caption{\label{fig:hs_M}}
    \end{subfigure}
    \begin{subfigure}[b]{0.495\textwidth}
        \centering
        \includegraphics[width=\linewidth]{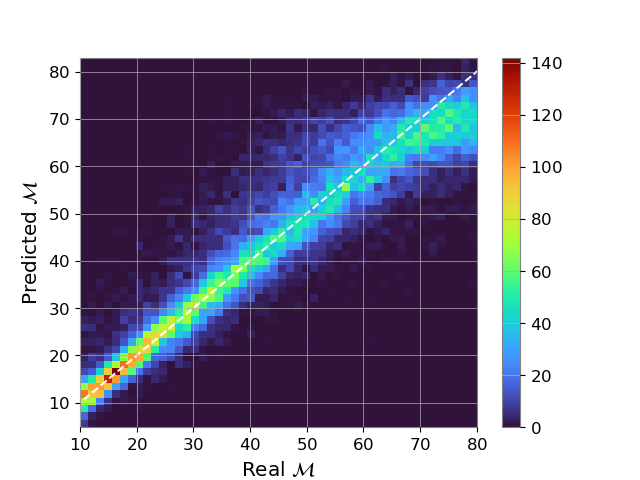}
        \caption{\label{fig:IMRHM_M}}
    \end{subfigure}
    \begin{subfigure}[b]{0.495\textwidth}
        \centering
        \includegraphics[width=\linewidth]{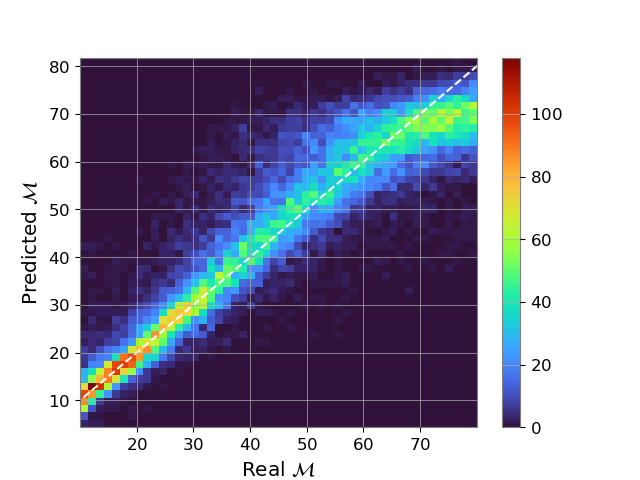}
        \caption{\label{fig:EOB_M}}
    \end{subfigure}
    \caption{Visual performance comparison for the estimation of the chirp mass $\mathcal{M}$ with all four approximants. Units are in $M_\odot$. \textbf{(a):} \texttt{IMRPhenomPv3}, \textbf{(b):} \texttt{NRHybSur3dq8}, \textbf{(c):} \texttt{IMRPhenomPv3HM}, and \textbf{(d):} \texttt{SEOBNRv4P}.}
    \label{fig:apcomp}
\end{figure*}
Settling on the top-performing TCN architecture, we move on to testing the effect of approximant selection on feature extraction from BBH merger waveforms. To do so, we follow a similar training procedure to the one detailed above but this time training 10 networks with the same TCN architecture for each of the datasets generated with the different approximants. Since NN training is a stochastic process at its core, training a network for each scenario multiple times allows us to build some statistics and analyse trends and fluctuations in the results, rather than focusing on a single scenario where a network converged. We limit the training to 40 epochs, as the results of the previous section suggest gains after this point are not too significant for the computation time spent. 

The evolution of the training and validation losses during the networks' training process is illustrated in \cref{fig:avg_train}, with the mean of the ten runs for each epoch being represented by the solid line, and the standard deviation being shown by the shaded area. Here we see that while the validation loss curves for \texttt{NRHybSur3dq8} and \texttt{IMRPhenomPv3} have significant overlap, there is a  slight deviation towards higher losses on the \texttt{SEOBNRv4P} run, and a large deviation towards lower losses for the \texttt{IMRPhenomPv3HM} scenario. In addition, the standard deviation band is significantly wider for the runs on the latter dataset.

After training we evaluate each network on its validation set, obtaining a root mean square error for each parameter $y^i$,
\begin{eqnarray}
\mathrm{RMSE}^j(x_\mathrm{in},y_\mathrm{truth},\Theta) = \sqrt{\frac{\sum_i [(y_\mathrm{truth}^i)^j-f(x_\mathrm{in}^i, \Theta)^j ]^2}{N}},
\nonumber \\
\end{eqnarray}
where $N$ is the number of elements evaluated, index $j$ refers to one of the parameters and $i$ is the index of the sample. To better analyse the distribution of these values for each of the approximants used, we show in \cref{fig:boxplots}  a collection of box plots. In these plots, each data point  (black markers) represents the RMSE value for the validation loss of an independent iteration of a TCN network. The horizontal line on each of the boxes shows the median value of the error, while the region encompassed between the median and the edge of the box represents the second and third quartiles of the data. The whiskers of the boxes extend up to $1.5$ times the inter-quartile range. Points not in between a box's whiskers are beyond the whisker range and are considered to be outliers. 

A first inspection of \cref{fig:boxplots} shows that while there are visible differences in the RMSE for all parameters depending on the chosen approximant, the range of these variations is not too large. In particular, the standard deviation of the median values for each of the ${\cal M},\,q,\,d_L$ and $\chi_\mathrm{eff}$ parameters are respectively, $1.3\times10^{-1}~\mathrm{M}_\odot,\,6.7\times10^{-3},\,29~\mathrm{Mpc},$ and $5.7\times10^{-3}$. We observe that TCN networks trained on \texttt{IMRPhenomPv3HM}  show a very significant advantage in recovering the luminosity distance of the injection. It is quite unclear as to why this difference should manifest. \texttt{IMRPhenomPv3HM}  also shows a smaller advantage in recovering the mass ratio $q$, while the chirp mass and effective spin are better recovered in the \texttt{IMRPhenomPv3} experiment. TCN instances trained on the \texttt{SEOBNRv4P} dataset stand out by having slightly worse parameter recovery than the other experiments, save for the case of the effective spin where it very slightly outperforms the \texttt{IMRPhenomPv3HM}  and \texttt{NRHybSur3dq8} cases.   

It is interesting to note that, despite what could have been expected, the presence of higher modes in an approximant (the case of \texttt{IMRPhenomPv3HM}  and \texttt{NRHybSur3dq8}) does not seem to systematically affect the ability of the TCN network to recover the injection parameters. This indicates that the increase in complexity in the data does not necessarily make it harder for the network to converge under the tested conditions.

\cref{fig:apcomp} shows an example of a visual comparison between the performances of networks trained on each approximant on the chirp mass parameter. In these 2D diagrams, obtained on the validation dataset for each approximant, the $x$-axis represents the injected values of the parameter, defined at generation time for each of the elements of the dataset, while the $y$-axis represents the predictions of each network. The colour in each bin shows the amount of events that fit within, translated into numbers by the colour bar on the right side of each plot. We observe that the network can infer the chirp mass values with reasonable accuracy, as expected from the previous discussion and \cref{fig:boxplots}. This kind of performance on one of the most dominant parameters is further indication that the main features of the gravitational information within the waveform are being successfully identified.

\begin{figure}
    \centering
    \begin{tikzpicture}    
        \node (img) {\includegraphics[width=\linewidth]{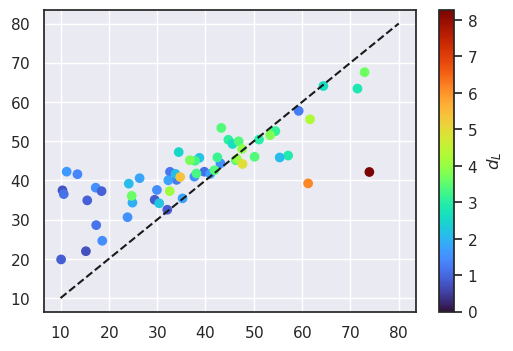}};
        \node[below=of img, node distance=0cm, yshift=1.2cm, xshift=-0.25cm] {Tabulated $\mathcal{M}$};
        \node[left=of img, node distance=0cm, rotate=90, anchor=center,yshift=-0.9cm] {Predicted $\mathcal{M}$};
    \end{tikzpicture}
    \caption{\label{fig:realdata} Prediction behaviour for the chirp mass as a function of $d_L$, for the studied LVK detections.
    }
\end{figure}

\subsection{Feature extraction from real data}

As a way to test the feature extraction capabilities of the studied method under real-world conditions, we apply our best performing network on actual GW events identified by the LVK collaboration. Given the parameter ranges of our training dataset, and the noise conditions of the training set, we limited this analysis to events from the O3 catalog with $10<\mathcal{M}<80~\mathrm{M}_\odot$ . The results for the recovery of the chirp mass $\mathcal{M}$ are plotted in \cref{fig:realdata}. Here we notice that for the considered events most predictions are within $10~\mathrm{M}_\odot$. Two events with luminosity distances beyond $6~\mathrm{Mpc}$ have network predictions significantly below the tabulated values, and there is a population of 7 low-mass events that is being assigned significantly higher masses by the network. Given the fact that the noise background varies significantly both between the analysed events, as well as between the events and the training set, it is encouraging to see that feature extracting performance is overall functional under different noise conditions and may be effective on real data. Furthermore, it is likely that training under more varied noise conditions and a finer covering of the parameter space in the training dataset could improve the generalization capacity of the network.

\section{Conclusion}

The existence of DL methods for time-series analysis, both for classification and forecasting, provides the GW community with the option to build data analysis pipelines upon a large body of work. In this paper we have applied a sample of neural networks, both convolutional and recurrent, to the estimation of physical parameters from GW signals produced in binary black hole mergers. Our specific goal has been to address if the choice of a particular waveform model or approximant for the creation of a dataset may condition the feature extraction ability of a trained network. To this aim we have used waveform datasets based on four of the most common waveform models available in the literature, \texttt{IMRPhenomPv3}, \texttt{IMRPhenomPv3HM}, \texttt{SEOBNRv4P}, and \texttt{NRHybSur3dq8}, injected into detector noise of the Advanced LIGO and Advanced Virgo detectors. Our study has been performed using the out-of-the-box network architectures provided by \texttt{tsai}~\cite{oguiza_tsai_2022}  with no further modifications thereof.

For feature extraction, we have found a significant advantage in performance for residual-based architectures, obtaining the best results with the Temporal Convolutional Network (TCN) architecture in particular, in terms of training and validation losses and absence of data overfitting. Using this network we have explored the effects of the choice of waveform approximant on parameter estimation. Differences in the performance of several TCN networks as a function of the waveform model have been found. In particular, TCN networks perform better on datasets generated with the \texttt{IMPhenomPv3HM} approximant. Interestingly, the presence of higher modes (and thus more complex waveform morphologies) did not seem to affect the results in a systematic way. Our best-performing network has also been applied on actual GW signals identified by the LVK collaboration. Our results show that the feature extracting performance of the network is reasonably robust against noise conditions and may be effective on real data. 

We conclude that, under the specific setup we have employed, based on out-of-the-box network architectures, the CBC approximant choice when creating a GW dataset impacts the feature extraction performance of a neural network. This means care should be taken when building a synthetic GW dataset for the training of neural networks, as test performance does not necessarily translate to data which is more faithful to NR simulations. A similar study using Bayesian machine-learning methods for parameter estimation, followed by a comparison to traditional Monte-Carlo samplers, might shed further light on the findings reported in this paper.

\begin{acknowledgments}

%
We warmly thank Felipe Freitas for many fruitful discussions during the course of this work and Melissa L\'opez for her useful comments.
OGF is supported by an FCT doctoral scholarship (reference UI/BD/154358/2022).
JCB is supported by a fellowship from ``la Caixa'' Foundation (ID
100010434) and from the European Union’s Horizon 2020 research and innovation programme under the Marie Skłodowska-Curie grant agreement No 847648. The fellowship code is LCF/BQ/PI20/11760016. JCB is also supported by the research grant PID2020-118635GB-I00 from the Spain-Ministerio de Ciencia e Innovaci\'{o}n.
JAF and ATF are supported by the Spanish Agencia Estatal de Investigaci\'on (grant PID2021-125485NB-C21) funded by MCIN/AEI/10.13039/501100011033 and ERDF A way of making Europe. 
Further support is provided by the Generalitat Valenciana (grant CIPROM/2022/49), by the EU's Horizon 2020 research and innovation (RISE) programme H2020-MSCA-RISE-2017 (FunFiCO-777740), and  by  the  European Horizon  Europe  staff  exchange  (SE)  programme HORIZON-MSCA-2021-SE-01 (NewFunFiCO-101086251).
AO is partially supported by FCT, under the Contract CERN/FIS-PAR/0037/2021.
Computations have been performed at the Artemisa cluster (UV-CSIC) co-funded by the European Union through the 2014-2020 FEDER Operative Programme of Comunitat Valenciana, project IDIFEDER/2018/048.
This material is based upon work supported by NSF's LIGO Laboratory which is a major facility fully funded by the National Science Foundation.

\end{acknowledgments}


\bibliographystyle{apsrev}
\bibliography{references}

\begin{thebibliography}{66}
\expandafter\ifx\csname natexlab\endcsname\relax\def\natexlab#1{#1}\fi
\expandafter\ifx\csname bibnamefont\endcsname\relax
  \def\bibnamefont#1{#1}\fi
\expandafter\ifx\csname bibfnamefont\endcsname\relax
  \def\bibfnamefont#1{#1}\fi
\expandafter\ifx\csname citenamefont\endcsname\relax
  \def\citenamefont#1{#1}\fi
\expandafter\ifx\csname url\endcsname\relax
  \def\url#1{\texttt{#1}}\fi
\expandafter\ifx\csname urlprefix\endcsname\relax\def\urlprefix{URL }\fi
\providecommand{\bibinfo}[2]{#2}
\providecommand{\eprint}[2][]{\url{#2}}

\bibitem[{\citenamefont{Abbott et~al.}(2019)}]{abbott_gwtc-1_2019}
\bibinfo{author}{\bibfnamefont{B.~P.} \bibnamefont{Abbott}}
  \bibnamefont{et~al.}, \bibinfo{journal}{Physical Review X}
  \textbf{\bibinfo{volume}{9}}, \bibinfo{pages}{031040} (\bibinfo{year}{2019}),
  \bibinfo{note}{1811.12907}.

\bibitem[{\citenamefont{Abbott
  et~al.}(2021{\natexlab{a}})}]{abbott_gwtc-2_2020}
\bibinfo{author}{\bibfnamefont{R.}~\bibnamefont{Abbott}} \bibnamefont{et~al.},
  \bibinfo{journal}{Phys. Rev. X} \textbf{\bibinfo{volume}{11}},
  \bibinfo{pages}{021053} (\bibinfo{year}{2021}{\natexlab{a}}),
  \eprint{2010.14527}.

\bibitem[{\citenamefont{Abbott
  et~al.}(2021{\natexlab{b}})}]{LIGOScientific:2021usb}
\bibinfo{author}{\bibfnamefont{R.}~\bibnamefont{Abbott}} \bibnamefont{et~al.}
  (\bibinfo{year}{2021}{\natexlab{b}}), \eprint{2108.01045}.

\bibitem[{\citenamefont{Abbott
  et~al.}(2021{\natexlab{c}})}]{the_ligo_scientific_collaboration_gwtc-3_2021}
\bibinfo{author}{\bibfnamefont{R.}~\bibnamefont{Abbott}} \bibnamefont{et~al.}
  (\bibinfo{year}{2021}{\natexlab{c}}), \eprint{2111.03606}.

\bibitem[{\citenamefont{Abbott et~al.}(2017)}]{TheLIGOScientific:2017qsa}
\bibinfo{author}{\bibfnamefont{B.~P.} \bibnamefont{Abbott}}
  \bibnamefont{et~al.}, \bibinfo{journal}{Phys. Rev. Lett.}
  \textbf{\bibinfo{volume}{119}}, \bibinfo{pages}{161101}
  (\bibinfo{year}{2017}), \eprint{1710.05832}.

\bibitem[{\citenamefont{Abbott et~al.}(2020)}]{GW190425}
\bibinfo{author}{\bibfnamefont{B.~P.} \bibnamefont{Abbott}}
  \bibnamefont{et~al.}, \bibinfo{journal}{\apjl}
  \textbf{\bibinfo{volume}{892}}, \bibinfo{eid}{L3} (\bibinfo{year}{2020}),
  \eprint{2001.01761}.

\bibitem[{\citenamefont{Abbott et~al.}(2021{\natexlab{d}})}]{NSBHs}
\bibinfo{author}{\bibfnamefont{R.}~\bibnamefont{Abbott}} \bibnamefont{et~al.},
  \bibinfo{journal}{Astrophys. J. Lett.} \textbf{\bibinfo{volume}{915}},
  \bibinfo{pages}{L5} (\bibinfo{year}{2021}{\natexlab{d}}),
  \eprint{2106.15163},
  \urlprefix\url{https://doi.org/10.3847/2041-8213/ac082e}.

\bibitem[{\citenamefont{Wainstein and Zubakov}(1962)}]{MatchedFilter}
\bibinfo{author}{\bibfnamefont{L.}~\bibnamefont{Wainstein}} \bibnamefont{and}
  \bibinfo{author}{\bibfnamefont{V.}~\bibnamefont{Zubakov}},
  \bibinfo{journal}{Prentice-Hall, Englewood Cliffs}  (\bibinfo{year}{1962}).

\bibitem[{\citenamefont{Allen et~al.}(2012)\citenamefont{Allen, Anderson,
  Brady, Brown, and Creighton}}]{FindChirp}
\bibinfo{author}{\bibfnamefont{B.}~\bibnamefont{Allen}},
  \bibinfo{author}{\bibfnamefont{W.~G.} \bibnamefont{Anderson}},
  \bibinfo{author}{\bibfnamefont{P.~R.} \bibnamefont{Brady}},
  \bibinfo{author}{\bibfnamefont{D.~A.} \bibnamefont{Brown}}, \bibnamefont{and}
  \bibinfo{author}{\bibfnamefont{J.~D.~E.} \bibnamefont{Creighton}},
  \bibinfo{journal}{Physical Review D} \textbf{\bibinfo{volume}{85}}
  (\bibinfo{year}{2012}),
  \urlprefix\url{https://doi.org/10.1103/physrevd.85.122006}.

\bibitem[{\citenamefont{Usman et~al.}(2016{\natexlab{a}})}]{Usman:2015kfa}
\bibinfo{author}{\bibfnamefont{S.~A.} \bibnamefont{Usman}}
  \bibnamefont{et~al.}, \bibinfo{journal}{Class. Quant. Grav.}
  \textbf{\bibinfo{volume}{33}}, \bibinfo{pages}{215004}
  (\bibinfo{year}{2016}{\natexlab{a}}), \eprint{1508.02357}.

\bibitem[{\citenamefont{Veitch et~al.}(2015)}]{Veitch:2014wba}
\bibinfo{author}{\bibfnamefont{J.}~\bibnamefont{Veitch}} \bibnamefont{et~al.},
  \bibinfo{journal}{Phys. Rev. D} \textbf{\bibinfo{volume}{91}},
  \bibinfo{pages}{042003} (\bibinfo{year}{2015}), \eprint{1409.7215}.

\bibitem[{\citenamefont{Ashton et~al.}(2019)}]{Ashton:2018jfp}
\bibinfo{author}{\bibfnamefont{G.}~\bibnamefont{Ashton}} \bibnamefont{et~al.},
  \bibinfo{journal}{Astrophys. J. Suppl.} \textbf{\bibinfo{volume}{241}},
  \bibinfo{pages}{27} (\bibinfo{year}{2019}), \eprint{1811.02042}.

\bibitem[{\citenamefont{Khan et~al.}(2016)\citenamefont{Khan, Husa, Hannam,
  Ohme, Pürrer, Forteza, and Bohé}}]{khan_frequency-domain_2016}
\bibinfo{author}{\bibfnamefont{S.}~\bibnamefont{Khan}},
  \bibinfo{author}{\bibfnamefont{S.}~\bibnamefont{Husa}},
  \bibinfo{author}{\bibfnamefont{M.}~\bibnamefont{Hannam}},
  \bibinfo{author}{\bibfnamefont{F.}~\bibnamefont{Ohme}},
  \bibinfo{author}{\bibfnamefont{M.}~\bibnamefont{Pürrer}},
  \bibinfo{author}{\bibfnamefont{X.~J.} \bibnamefont{Forteza}},
  \bibnamefont{and} \bibinfo{author}{\bibfnamefont{A.}~\bibnamefont{Bohé}},
  \bibinfo{journal}{Phys. Rev. D} \textbf{\bibinfo{volume}{93}},
  \bibinfo{pages}{044007} (\bibinfo{year}{2016}), ISSN
  \bibinfo{issn}{2470-0010, 2470-0029}, \bibinfo{note}{arXiv: 1508.07253},
  \urlprefix\url{http://arxiv.org/abs/1508.07253}.

\bibitem[{\citenamefont{Khan et~al.}(2019)\citenamefont{Khan, Chatziioannou,
  Hannam, and Ohme}}]{khan_phenomenological_2019}
\bibinfo{author}{\bibfnamefont{S.}~\bibnamefont{Khan}},
  \bibinfo{author}{\bibfnamefont{K.}~\bibnamefont{Chatziioannou}},
  \bibinfo{author}{\bibfnamefont{M.}~\bibnamefont{Hannam}}, \bibnamefont{and}
  \bibinfo{author}{\bibfnamefont{F.}~\bibnamefont{Ohme}},
  \bibinfo{journal}{Phys. Rev. D} \textbf{\bibinfo{volume}{100}},
  \bibinfo{pages}{024059} (\bibinfo{year}{2019}), ISSN
  \bibinfo{issn}{2470-0010, 2470-0029}, \bibinfo{note}{arXiv:1809.10113
  [gr-qc]}, \urlprefix\url{http://arxiv.org/abs/1809.10113}.

\bibitem[{\citenamefont{Khan et~al.}(2020)\citenamefont{Khan, Ohme,
  Chatziioannou, and Hannam}}]{khan_including_2020}
\bibinfo{author}{\bibfnamefont{S.}~\bibnamefont{Khan}},
  \bibinfo{author}{\bibfnamefont{F.}~\bibnamefont{Ohme}},
  \bibinfo{author}{\bibfnamefont{K.}~\bibnamefont{Chatziioannou}},
  \bibnamefont{and} \bibinfo{author}{\bibfnamefont{M.}~\bibnamefont{Hannam}},
  \bibinfo{journal}{Physical Review D} \textbf{\bibinfo{volume}{101}}
  (\bibinfo{year}{2020}), ISSN \bibinfo{issn}{2470-0010},
  \bibinfo{note}{publisher: American Physical Society (APS)},
  \urlprefix\url{https://dx.doi.org/10.1103/physrevd.101.024056}.

\bibitem[{\citenamefont{Varma et~al.}(2019)\citenamefont{Varma, Field, Scheel,
  Blackman, Kidder, and Pfeiffer}}]{varma_surrogate_2019}
\bibinfo{author}{\bibfnamefont{V.}~\bibnamefont{Varma}},
  \bibinfo{author}{\bibfnamefont{S.~E.} \bibnamefont{Field}},
  \bibinfo{author}{\bibfnamefont{M.~A.} \bibnamefont{Scheel}},
  \bibinfo{author}{\bibfnamefont{J.}~\bibnamefont{Blackman}},
  \bibinfo{author}{\bibfnamefont{L.~E.} \bibnamefont{Kidder}},
  \bibnamefont{and} \bibinfo{author}{\bibfnamefont{H.~P.}
  \bibnamefont{Pfeiffer}}, \bibinfo{journal}{Phys. Rev. D}
  \textbf{\bibinfo{volume}{99}}, \bibinfo{pages}{064045}
  (\bibinfo{year}{2019}), ISSN \bibinfo{issn}{2470-0010, 2470-0029},
  \bibinfo{note}{arXiv:1812.07865 [gr-qc]},
  \urlprefix\url{http://arxiv.org/abs/1812.07865}.

\bibitem[{\citenamefont{Bohé et~al.}(2017)\citenamefont{Bohé, Shao,
  Taracchini, Buonanno, Babak, Harry, Hinder, Ossokine, Pürrer, Raymond
  et~al.}}]{bohe_improved_2017}
\bibinfo{author}{\bibfnamefont{A.}~\bibnamefont{Bohé}},
  \bibinfo{author}{\bibfnamefont{L.}~\bibnamefont{Shao}},
  \bibinfo{author}{\bibfnamefont{A.}~\bibnamefont{Taracchini}},
  \bibinfo{author}{\bibfnamefont{A.}~\bibnamefont{Buonanno}},
  \bibinfo{author}{\bibfnamefont{S.}~\bibnamefont{Babak}},
  \bibinfo{author}{\bibfnamefont{I.~W.} \bibnamefont{Harry}},
  \bibinfo{author}{\bibfnamefont{I.}~\bibnamefont{Hinder}},
  \bibinfo{author}{\bibfnamefont{S.}~\bibnamefont{Ossokine}},
  \bibinfo{author}{\bibfnamefont{M.}~\bibnamefont{Pürrer}},
  \bibinfo{author}{\bibfnamefont{V.}~\bibnamefont{Raymond}},
  \bibnamefont{et~al.}, \bibinfo{journal}{Phys. Rev. D}
  \textbf{\bibinfo{volume}{95}}, \bibinfo{pages}{044028}
  (\bibinfo{year}{2017}), \bibinfo{note}{publisher: American Physical Society},
  \urlprefix\url{https://link.aps.org/doi/10.1103/PhysRevD.95.044028}.

\bibitem[{\citenamefont{Cao and Han}(2017)}]{cao_waveform_2017}
\bibinfo{author}{\bibfnamefont{Z.}~\bibnamefont{Cao}} \bibnamefont{and}
  \bibinfo{author}{\bibfnamefont{W.-B.} \bibnamefont{Han}},
  \bibinfo{journal}{Phys. Rev. D} \textbf{\bibinfo{volume}{96}},
  \bibinfo{pages}{044028} (\bibinfo{year}{2017}), ISSN
  \bibinfo{issn}{2470-0010, 2470-0029},
  \urlprefix\url{https://link.aps.org/doi/10.1103/PhysRevD.96.044028}.

\bibitem[{\citenamefont{Akcay et~al.}(2020)\citenamefont{Akcay, Gamba, and
  Bernuzzi}}]{akcay_hybrid_2020}
\bibinfo{author}{\bibfnamefont{S.}~\bibnamefont{Akcay}},
  \bibinfo{author}{\bibfnamefont{R.}~\bibnamefont{Gamba}}, \bibnamefont{and}
  \bibinfo{author}{\bibfnamefont{S.}~\bibnamefont{Bernuzzi}},
  \bibinfo{journal}{arXiv:2005.05338 [gr-qc]}  (\bibinfo{year}{2020}),
  \bibinfo{note}{arXiv: 2005.05338},
  \urlprefix\url{http://arxiv.org/abs/2005.05338}.

\bibitem[{\citenamefont{Nagar et~al.}(2018)\citenamefont{Nagar, Bernuzzi,
  Del~Pozzo, Riemenschneider, Akcay, Carullo, Fleig, Babak, Tsang, Colleoni
  et~al.}}]{nagar_time-domain_2018}
\bibinfo{author}{\bibfnamefont{A.}~\bibnamefont{Nagar}},
  \bibinfo{author}{\bibfnamefont{S.}~\bibnamefont{Bernuzzi}},
  \bibinfo{author}{\bibfnamefont{W.}~\bibnamefont{Del~Pozzo}},
  \bibinfo{author}{\bibfnamefont{G.}~\bibnamefont{Riemenschneider}},
  \bibinfo{author}{\bibfnamefont{S.}~\bibnamefont{Akcay}},
  \bibinfo{author}{\bibfnamefont{G.}~\bibnamefont{Carullo}},
  \bibinfo{author}{\bibfnamefont{P.}~\bibnamefont{Fleig}},
  \bibinfo{author}{\bibfnamefont{S.}~\bibnamefont{Babak}},
  \bibinfo{author}{\bibfnamefont{K.~W.} \bibnamefont{Tsang}},
  \bibinfo{author}{\bibfnamefont{M.}~\bibnamefont{Colleoni}},
  \bibnamefont{et~al.}, \bibinfo{journal}{{\textbackslash}prd}
  \textbf{\bibinfo{volume}{98}}, \bibinfo{pages}{104052}
  (\bibinfo{year}{2018}), \bibinfo{note}{1806.01772}.

\bibitem[{\citenamefont{Belgacem et~al.}(2019)\citenamefont{Belgacem, Dirian,
  Foffa, Howell, Maggiore, and Regimbau}}]{belgacem_cosmology_2019}
\bibinfo{author}{\bibfnamefont{E.}~\bibnamefont{Belgacem}},
  \bibinfo{author}{\bibfnamefont{Y.}~\bibnamefont{Dirian}},
  \bibinfo{author}{\bibfnamefont{S.}~\bibnamefont{Foffa}},
  \bibinfo{author}{\bibfnamefont{E.~J.} \bibnamefont{Howell}},
  \bibinfo{author}{\bibfnamefont{M.}~\bibnamefont{Maggiore}}, \bibnamefont{and}
  \bibinfo{author}{\bibfnamefont{T.}~\bibnamefont{Regimbau}},
  \bibinfo{journal}{Journal of Cosmology and Astroparticle Physics}
  \textbf{\bibinfo{volume}{2019}}, \bibinfo{pages}{015} (\bibinfo{year}{2019}),
  \bibinfo{note}{publisher: IOP Publishing},
  \urlprefix\url{https://doi.org/10.1088%2F1475-7516%2F2019%2F08%2F015}.

\bibitem[{\citenamefont{{Green} et~al.}(2020)\citenamefont{{Green}, {Simpson},
  and {Gair}}}]{Green:2020}
\bibinfo{author}{\bibfnamefont{S.~R.} \bibnamefont{{Green}}},
  \bibinfo{author}{\bibfnamefont{C.}~\bibnamefont{{Simpson}}},
  \bibnamefont{and} \bibinfo{author}{\bibfnamefont{J.}~\bibnamefont{{Gair}}},
  \bibinfo{journal}{\prd} \textbf{\bibinfo{volume}{102}}, \bibinfo{eid}{104057}
  (\bibinfo{year}{2020}), \eprint{2002.07656}.

\bibitem[{\citenamefont{Dax et~al.}(2021)\citenamefont{Dax, Green, Gair, Macke,
  Buonanno, and Schölkopf}}]{dax_real-time_2021}
\bibinfo{author}{\bibfnamefont{M.}~\bibnamefont{Dax}},
  \bibinfo{author}{\bibfnamefont{S.~R.} \bibnamefont{Green}},
  \bibinfo{author}{\bibfnamefont{J.}~\bibnamefont{Gair}},
  \bibinfo{author}{\bibfnamefont{J.~H.} \bibnamefont{Macke}},
  \bibinfo{author}{\bibfnamefont{A.}~\bibnamefont{Buonanno}}, \bibnamefont{and}
  \bibinfo{author}{\bibfnamefont{B.}~\bibnamefont{Schölkopf}},
  \bibinfo{journal}{arXiv:2106.12594 [astro-ph, physics:gr-qc]}
  (\bibinfo{year}{2021}), \bibinfo{note}{arXiv: 2106.12594},
  \urlprefix\url{http://arxiv.org/abs/2106.12594}.

\bibitem[{\citenamefont{Williams et~al.}(2021)\citenamefont{Williams, Veitch,
  and Messenger}}]{williams_nested_2021}
\bibinfo{author}{\bibfnamefont{M.~J.} \bibnamefont{Williams}},
  \bibinfo{author}{\bibfnamefont{J.}~\bibnamefont{Veitch}}, \bibnamefont{and}
  \bibinfo{author}{\bibfnamefont{C.}~\bibnamefont{Messenger}},
  \bibinfo{journal}{Phys. Rev. D} \textbf{\bibinfo{volume}{103}},
  \bibinfo{pages}{103006} (\bibinfo{year}{2021}), ISSN
  \bibinfo{issn}{2470-0010, 2470-0029}, \bibinfo{note}{arXiv:2102.11056
  [astro-ph, physics:gr-qc]}, \urlprefix\url{http://arxiv.org/abs/2102.11056}.

\bibitem[{\citenamefont{Bayley et~al.}(2022)\citenamefont{Bayley, Messenger,
  and Woan}}]{bayley_rapid_2022}
\bibinfo{author}{\bibfnamefont{J.}~\bibnamefont{Bayley}},
  \bibinfo{author}{\bibfnamefont{C.}~\bibnamefont{Messenger}},
  \bibnamefont{and} \bibinfo{author}{\bibfnamefont{G.}~\bibnamefont{Woan}},
  \emph{\bibinfo{title}{Rapid parameter estimation for an all-sky continuous
  gravitational wave search using conditional varitational auto-encoders}}
  (\bibinfo{year}{2022}), \bibinfo{note}{arXiv:2209.02031 [astro-ph]},
  \urlprefix\url{http://arxiv.org/abs/2209.02031}.

\bibitem[{\citenamefont{{Gabbard} et~al.}(2022)\citenamefont{{Gabbard},
  {Messenger}, {Heng}, {Tonolini}, and {Murray-Smith}}}]{Gabbard:2022}
\bibinfo{author}{\bibfnamefont{H.}~\bibnamefont{{Gabbard}}},
  \bibinfo{author}{\bibfnamefont{C.}~\bibnamefont{{Messenger}}},
  \bibinfo{author}{\bibfnamefont{I.~S.} \bibnamefont{{Heng}}},
  \bibinfo{author}{\bibfnamefont{F.}~\bibnamefont{{Tonolini}}},
  \bibnamefont{and}
  \bibinfo{author}{\bibfnamefont{R.}~\bibnamefont{{Murray-Smith}}},
  \bibinfo{journal}{Nature Physics} \textbf{\bibinfo{volume}{18}},
  \bibinfo{pages}{112} (\bibinfo{year}{2022}), \eprint{1909.06296}.

\bibitem[{\citenamefont{{Dax} et~al.}(2023)\citenamefont{{Dax}, {Green},
  {Gair}, {P{\"u}rrer}, {Wildberger}, {Macke}, {Buonanno}, and
  {Sch{\"o}lkopf}}}]{Dax:2023}
\bibinfo{author}{\bibfnamefont{M.}~\bibnamefont{{Dax}}},
  \bibinfo{author}{\bibfnamefont{S.~R.} \bibnamefont{{Green}}},
  \bibinfo{author}{\bibfnamefont{J.}~\bibnamefont{{Gair}}},
  \bibinfo{author}{\bibfnamefont{M.}~\bibnamefont{{P{\"u}rrer}}},
  \bibinfo{author}{\bibfnamefont{J.}~\bibnamefont{{Wildberger}}},
  \bibinfo{author}{\bibfnamefont{J.~H.} \bibnamefont{{Macke}}},
  \bibinfo{author}{\bibfnamefont{A.}~\bibnamefont{{Buonanno}}},
  \bibnamefont{and}
  \bibinfo{author}{\bibfnamefont{B.}~\bibnamefont{{Sch{\"o}lkopf}}},
  \bibinfo{journal}{\prl} \textbf{\bibinfo{volume}{130}}, \bibinfo{eid}{171403}
  (\bibinfo{year}{2023}), \eprint{2210.05686}.

\bibitem[{\citenamefont{Bhardwaj et~al.}(2023)\citenamefont{Bhardwaj, Alvey,
  Miller, Nissanke, and Weniger}}]{bhardwaj2023peregrine}
\bibinfo{author}{\bibfnamefont{U.}~\bibnamefont{Bhardwaj}},
  \bibinfo{author}{\bibfnamefont{J.}~\bibnamefont{Alvey}},
  \bibinfo{author}{\bibfnamefont{B.~K.} \bibnamefont{Miller}},
  \bibinfo{author}{\bibfnamefont{S.}~\bibnamefont{Nissanke}}, \bibnamefont{and}
  \bibinfo{author}{\bibfnamefont{C.}~\bibnamefont{Weniger}},
  \emph{\bibinfo{title}{Peregrine: Sequential simulation-based inference for
  gravitational wave signals}} (\bibinfo{year}{2023}), \eprint{2304.02035}.

\bibitem[{\citenamefont{{Huerta} et~al.}(2019)\citenamefont{{Huerta}, {Allen},
  {Andreoni}, {Antelis}, {Bachelet}, {Berriman}, {Bianco}, {Biswas}, {Carrasco
  Kind}, {Chard} et~al.}}]{Huerta:2019}
\bibinfo{author}{\bibfnamefont{E.~A.} \bibnamefont{{Huerta}}},
  \bibinfo{author}{\bibfnamefont{G.}~\bibnamefont{{Allen}}},
  \bibinfo{author}{\bibfnamefont{I.}~\bibnamefont{{Andreoni}}},
  \bibinfo{author}{\bibfnamefont{J.~M.} \bibnamefont{{Antelis}}},
  \bibinfo{author}{\bibfnamefont{E.}~\bibnamefont{{Bachelet}}},
  \bibinfo{author}{\bibfnamefont{G.~B.} \bibnamefont{{Berriman}}},
  \bibinfo{author}{\bibfnamefont{F.~B.} \bibnamefont{{Bianco}}},
  \bibinfo{author}{\bibfnamefont{R.}~\bibnamefont{{Biswas}}},
  \bibinfo{author}{\bibfnamefont{M.}~\bibnamefont{{Carrasco Kind}}},
  \bibinfo{author}{\bibfnamefont{K.}~\bibnamefont{{Chard}}},
  \bibnamefont{et~al.}, \bibinfo{journal}{Nature Reviews Physics}
  \textbf{\bibinfo{volume}{1}}, \bibinfo{pages}{600} (\bibinfo{year}{2019}),
  \eprint{1911.11779}.

\bibitem[{\citenamefont{{Cuoco} et~al.}(2020)\citenamefont{{Cuoco}, {Powell},
  {Cavagli{\`a}}, {Ackley}, {Bejger}, {Chatterjee}, {Coughlin}, {Coughlin},
  {Easter}, {Essick} et~al.}}]{Cuoco:2020}
\bibinfo{author}{\bibfnamefont{E.}~\bibnamefont{{Cuoco}}},
  \bibinfo{author}{\bibfnamefont{J.}~\bibnamefont{{Powell}}},
  \bibinfo{author}{\bibfnamefont{M.}~\bibnamefont{{Cavagli{\`a}}}},
  \bibinfo{author}{\bibfnamefont{K.}~\bibnamefont{{Ackley}}},
  \bibinfo{author}{\bibfnamefont{M.}~\bibnamefont{{Bejger}}},
  \bibinfo{author}{\bibfnamefont{C.}~\bibnamefont{{Chatterjee}}},
  \bibinfo{author}{\bibfnamefont{M.}~\bibnamefont{{Coughlin}}},
  \bibinfo{author}{\bibfnamefont{S.}~\bibnamefont{{Coughlin}}},
  \bibinfo{author}{\bibfnamefont{P.}~\bibnamefont{{Easter}}},
  \bibinfo{author}{\bibfnamefont{R.}~\bibnamefont{{Essick}}},
  \bibnamefont{et~al.}, \bibinfo{journal}{arXiv e-prints}
  \bibinfo{eid}{arXiv:2005.03745} (\bibinfo{year}{2020}), \eprint{2005.03745}.

\bibitem[{\citenamefont{Álvares et~al.}(2021)\citenamefont{Álvares, Font,
  Freitas, Freitas, Morais, Nunes, Onofre, and
  Torres-Forné}}]{alvares_exploring_2021}
\bibinfo{author}{\bibfnamefont{J.~D.} \bibnamefont{Álvares}},
  \bibinfo{author}{\bibfnamefont{J.~A.} \bibnamefont{Font}},
  \bibinfo{author}{\bibfnamefont{F.~F.} \bibnamefont{Freitas}},
  \bibinfo{author}{\bibfnamefont{O.~G.} \bibnamefont{Freitas}},
  \bibinfo{author}{\bibfnamefont{A.~P.} \bibnamefont{Morais}},
  \bibinfo{author}{\bibfnamefont{S.}~\bibnamefont{Nunes}},
  \bibinfo{author}{\bibfnamefont{A.}~\bibnamefont{Onofre}}, \bibnamefont{and}
  \bibinfo{author}{\bibfnamefont{A.}~\bibnamefont{Torres-Forné}},
  \bibinfo{journal}{Class. Quantum Grav.} \textbf{\bibinfo{volume}{38}},
  \bibinfo{pages}{155010} (\bibinfo{year}{2021}), ISSN
  \bibinfo{issn}{0264-9381, 1361-6382},
  \urlprefix\url{https://iopscience.iop.org/article/10.1088/1361-6382/ ac0455}.

\bibitem[{\citenamefont{Boudart and Fays}(2022)}]{ALBUS}
\bibinfo{author}{\bibfnamefont{V.}~\bibnamefont{Boudart}} \bibnamefont{and}
  \bibinfo{author}{\bibfnamefont{M.}~\bibnamefont{Fays}}, in
  \emph{\bibinfo{booktitle}{2022 IEEE International Conference on Big Data (Big
  Data)}} (\bibinfo{year}{2022}), pp. \bibinfo{pages}{6599--6601}.

\bibitem[{\citenamefont{Sch\"afer et~al.}(2023)\citenamefont{Sch\"afer,
  Zelenka, Nitz, Wang, Wu, Guo, Cao, Ren, Nousi, Stergioulas
  et~al.}}]{mockdatachal}
\bibinfo{author}{\bibfnamefont{M.~B.} \bibnamefont{Sch\"afer}},
  \bibinfo{author}{\bibfnamefont{O.~c.~v.} \bibnamefont{Zelenka}},
  \bibinfo{author}{\bibfnamefont{A.~H.} \bibnamefont{Nitz}},
  \bibinfo{author}{\bibfnamefont{H.}~\bibnamefont{Wang}},
  \bibinfo{author}{\bibfnamefont{S.}~\bibnamefont{Wu}},
  \bibinfo{author}{\bibfnamefont{Z.-K.} \bibnamefont{Guo}},
  \bibinfo{author}{\bibfnamefont{Z.}~\bibnamefont{Cao}},
  \bibinfo{author}{\bibfnamefont{Z.}~\bibnamefont{Ren}},
  \bibinfo{author}{\bibfnamefont{P.}~\bibnamefont{Nousi}},
  \bibinfo{author}{\bibfnamefont{N.}~\bibnamefont{Stergioulas}},
  \bibnamefont{et~al.}, \bibinfo{journal}{Phys. Rev. D}
  \textbf{\bibinfo{volume}{107}}, \bibinfo{pages}{023021}
  (\bibinfo{year}{2023}),
  \urlprefix\url{https://link.aps.org/doi/10.1103/PhysRevD.107.023021}.

\bibitem[{\citenamefont{Torres-Forné et~al.}(2016)\citenamefont{Torres-Forné,
  Marquina, Font, and Ibáñez}}]{torres-forne_denoising_2016}
\bibinfo{author}{\bibfnamefont{A.}~\bibnamefont{Torres-Forné}},
  \bibinfo{author}{\bibfnamefont{A.}~\bibnamefont{Marquina}},
  \bibinfo{author}{\bibfnamefont{J.~A.} \bibnamefont{Font}}, \bibnamefont{and}
  \bibinfo{author}{\bibfnamefont{J.~M.} \bibnamefont{Ibáñez}},
  \bibinfo{journal}{Phys. Rev. D} \textbf{\bibinfo{volume}{94}},
  \bibinfo{pages}{124040} (\bibinfo{year}{2016}), \bibinfo{note}{1612.01305}.

\bibitem[{\citenamefont{Torres-Forné et~al.}(2020)\citenamefont{Torres-Forné,
  Cuoco, Font, and Marquina}}]{torres-forne_application_2020}
\bibinfo{author}{\bibfnamefont{A.}~\bibnamefont{Torres-Forné}},
  \bibinfo{author}{\bibfnamefont{E.}~\bibnamefont{Cuoco}},
  \bibinfo{author}{\bibfnamefont{J.~A.} \bibnamefont{Font}}, \bibnamefont{and}
  \bibinfo{author}{\bibfnamefont{A.}~\bibnamefont{Marquina}},
  \bibinfo{journal}{Phys. Rev. D} \textbf{\bibinfo{volume}{102}},
  \bibinfo{pages}{023011} (\bibinfo{year}{2020}), \bibinfo{note}{publisher:
  American Physical Society},
  \urlprefix\url{https://link.aps.org/doi/10.1103/PhysRevD.102.023011}.

\bibitem[{\citenamefont{Schmidt et~al.}(2021)\citenamefont{Schmidt, Breschi,
  Gamba, Pagano, Rettegno, Riemenschneider, Bernuzzi, Nagar, and
  Del~Pozzo}}]{stefano_mlgw}
\bibinfo{author}{\bibfnamefont{S.}~\bibnamefont{Schmidt}},
  \bibinfo{author}{\bibfnamefont{M.}~\bibnamefont{Breschi}},
  \bibinfo{author}{\bibfnamefont{R.}~\bibnamefont{Gamba}},
  \bibinfo{author}{\bibfnamefont{G.}~\bibnamefont{Pagano}},
  \bibinfo{author}{\bibfnamefont{P.}~\bibnamefont{Rettegno}},
  \bibinfo{author}{\bibfnamefont{G.}~\bibnamefont{Riemenschneider}},
  \bibinfo{author}{\bibfnamefont{S.}~\bibnamefont{Bernuzzi}},
  \bibinfo{author}{\bibfnamefont{A.}~\bibnamefont{Nagar}}, \bibnamefont{and}
  \bibinfo{author}{\bibfnamefont{W.}~\bibnamefont{Del~Pozzo}},
  \bibinfo{journal}{Phys. Rev. D} \textbf{\bibinfo{volume}{103}},
  \bibinfo{pages}{043020} (\bibinfo{year}{2021}),
  \urlprefix\url{https://link.aps.org/doi/10.1103/PhysRevD.103.043020}.

\bibitem[{\citenamefont{Tissino et~al.}(2023)\citenamefont{Tissino, Carullo,
  Breschi, Gamba, Schmidt, and Bernuzzi}}]{stefano_bns}
\bibinfo{author}{\bibfnamefont{J.}~\bibnamefont{Tissino}},
  \bibinfo{author}{\bibfnamefont{G.}~\bibnamefont{Carullo}},
  \bibinfo{author}{\bibfnamefont{M.}~\bibnamefont{Breschi}},
  \bibinfo{author}{\bibfnamefont{R.}~\bibnamefont{Gamba}},
  \bibinfo{author}{\bibfnamefont{S.}~\bibnamefont{Schmidt}}, \bibnamefont{and}
  \bibinfo{author}{\bibfnamefont{S.}~\bibnamefont{Bernuzzi}},
  \bibinfo{journal}{Phys. Rev. D} \textbf{\bibinfo{volume}{107}},
  \bibinfo{pages}{084037} (\bibinfo{year}{2023}),
  \urlprefix\url{https://link.aps.org/doi/10.1103/PhysRevD.107.084037}.

\bibitem[{\citenamefont{Lopez et~al.}(2022{\natexlab{a}})\citenamefont{Lopez,
  Boudart, Buijsman, Reza, and Caudill}}]{melissa_gans}
\bibinfo{author}{\bibfnamefont{M.}~\bibnamefont{Lopez}},
  \bibinfo{author}{\bibfnamefont{V.}~\bibnamefont{Boudart}},
  \bibinfo{author}{\bibfnamefont{K.}~\bibnamefont{Buijsman}},
  \bibinfo{author}{\bibfnamefont{A.}~\bibnamefont{Reza}}, \bibnamefont{and}
  \bibinfo{author}{\bibfnamefont{S.}~\bibnamefont{Caudill}},
  \bibinfo{journal}{Phys. Rev. D} \textbf{\bibinfo{volume}{106}},
  \bibinfo{pages}{023027} (\bibinfo{year}{2022}{\natexlab{a}}),
  \urlprefix\url{https://link.aps.org/doi/10.1103/PhysRevD.106.023027}.

\bibitem[{\citenamefont{Lopez et~al.}(2022{\natexlab{b}})\citenamefont{Lopez,
  Boudart, Schmidt, and Caudill}}]{lopez_gengli}
\bibinfo{author}{\bibfnamefont{M.}~\bibnamefont{Lopez}},
  \bibinfo{author}{\bibfnamefont{V.}~\bibnamefont{Boudart}},
  \bibinfo{author}{\bibfnamefont{S.}~\bibnamefont{Schmidt}}, \bibnamefont{and}
  \bibinfo{author}{\bibfnamefont{S.}~\bibnamefont{Caudill}},
  \emph{\bibinfo{title}{Simulating transient noise bursts in ligo with gengli}}
  (\bibinfo{year}{2022}{\natexlab{b}}), \eprint{2205.09204}.

\bibitem[{\citenamefont{Fawaz et~al.}(2018)\citenamefont{Fawaz, Forestier,
  Weber, Idoumghar, and Muller}}]{fawaz_deep_2018}
\bibinfo{author}{\bibfnamefont{H.~I.} \bibnamefont{Fawaz}},
  \bibinfo{author}{\bibfnamefont{G.}~\bibnamefont{Forestier}},
  \bibinfo{author}{\bibfnamefont{J.}~\bibnamefont{Weber}},
  \bibinfo{author}{\bibfnamefont{L.}~\bibnamefont{Idoumghar}},
  \bibnamefont{and} \bibinfo{author}{\bibfnamefont{P.-A.} \bibnamefont{Muller}}
  (\bibinfo{year}{2018}), \bibinfo{note}{publisher: arXiv Version Number: 4},
  \urlprefix\url{https://arxiv.org/abs/1809.04356}.

\bibitem[{\citenamefont{Lara-Benítez et~al.}(2021)\citenamefont{Lara-Benítez,
  Carranza-García, and Riquelme}}]{lara-benitez_experimental_2021}
\bibinfo{author}{\bibfnamefont{P.}~\bibnamefont{Lara-Benítez}},
  \bibinfo{author}{\bibfnamefont{M.}~\bibnamefont{Carranza-García}},
  \bibnamefont{and} \bibinfo{author}{\bibfnamefont{J.~C.}
  \bibnamefont{Riquelme}}, \bibinfo{journal}{Int. J. Neur. Syst.}
  \textbf{\bibinfo{volume}{31}}, \bibinfo{pages}{2130001}
  (\bibinfo{year}{2021}), ISSN \bibinfo{issn}{0129-0657, 1793-6462},
  \bibinfo{note}{arXiv:2103.12057 [cs]},
  \urlprefix\url{http://arxiv.org/abs/2103.12057}.

\bibitem[{\citenamefont{Tan et~al.}(2021)\citenamefont{Tan, Bergmeir,
  Petitjean, and Webb}}]{tan_time_2021}
\bibinfo{author}{\bibfnamefont{C.~W.} \bibnamefont{Tan}},
  \bibinfo{author}{\bibfnamefont{C.}~\bibnamefont{Bergmeir}},
  \bibinfo{author}{\bibfnamefont{F.}~\bibnamefont{Petitjean}},
  \bibnamefont{and} \bibinfo{author}{\bibfnamefont{G.~I.} \bibnamefont{Webb}},
  \bibinfo{journal}{Data Mining and Knowledge Discovery}
  \textbf{\bibinfo{volume}{35}}, \bibinfo{pages}{1032} (\bibinfo{year}{2021}),
  ISSN \bibinfo{issn}{1384-5810}, \bibinfo{note}{publisher: Springer Science
  and Business Media LLC},
  \urlprefix\url{https://dx.doi.org/10.1007/s10618-021-00745-9}.

\bibitem[{\citenamefont{Santamaria et~al.}(2010)\citenamefont{Santamaria, Ohme,
  Ajith, Bruegmann, Dorband, Hannam, Husa, Moesta, Pollney, Reisswig
  et~al.}}]{santamaria_matching_2010}
\bibinfo{author}{\bibfnamefont{L.}~\bibnamefont{Santamaria}},
  \bibinfo{author}{\bibfnamefont{F.}~\bibnamefont{Ohme}},
  \bibinfo{author}{\bibfnamefont{P.}~\bibnamefont{Ajith}},
  \bibinfo{author}{\bibfnamefont{B.}~\bibnamefont{Bruegmann}},
  \bibinfo{author}{\bibfnamefont{N.}~\bibnamefont{Dorband}},
  \bibinfo{author}{\bibfnamefont{M.}~\bibnamefont{Hannam}},
  \bibinfo{author}{\bibfnamefont{S.}~\bibnamefont{Husa}},
  \bibinfo{author}{\bibfnamefont{P.}~\bibnamefont{Moesta}},
  \bibinfo{author}{\bibfnamefont{D.}~\bibnamefont{Pollney}},
  \bibinfo{author}{\bibfnamefont{C.}~\bibnamefont{Reisswig}},
  \bibnamefont{et~al.}, \bibinfo{journal}{Phys. Rev. D}
  \textbf{\bibinfo{volume}{82}}, \bibinfo{pages}{064016}
  (\bibinfo{year}{2010}), ISSN \bibinfo{issn}{1550-7998, 1550-2368},
  \bibinfo{note}{arXiv: 1005.3306},
  \urlprefix\url{http://arxiv.org/abs/1005.3306}.

\bibitem[{\citenamefont{Cotesta et~al.}(2020)\citenamefont{Cotesta, Marsat, and
  Pürrer}}]{cotesta_frequency_2020}
\bibinfo{author}{\bibfnamefont{R.}~\bibnamefont{Cotesta}},
  \bibinfo{author}{\bibfnamefont{S.}~\bibnamefont{Marsat}}, \bibnamefont{and}
  \bibinfo{author}{\bibfnamefont{M.}~\bibnamefont{Pürrer}},
  \bibinfo{journal}{Phys. Rev. D} \textbf{\bibinfo{volume}{101}},
  \bibinfo{pages}{124040} (\bibinfo{year}{2020}), \bibinfo{note}{2003.12079}.

\bibitem[{\citenamefont{Oguiza}(2022)}]{oguiza_tsai_2022}
\bibinfo{author}{\bibfnamefont{I.}~\bibnamefont{Oguiza}},
  \emph{\bibinfo{title}{tsai - {A} state-of-the-art deep learning library for
  time series and sequential data}} (\bibinfo{year}{2022}),
  \urlprefix\url{https://github.com/timeseriesAI/tsai}.

\bibitem[{\citenamefont{{Schmidt} et~al.}(2015)\citenamefont{{Schmidt}, {Ohme},
  and {Hannam}}}]{Schmidt:2015}
\bibinfo{author}{\bibfnamefont{P.}~\bibnamefont{{Schmidt}}},
  \bibinfo{author}{\bibfnamefont{F.}~\bibnamefont{{Ohme}}}, \bibnamefont{and}
  \bibinfo{author}{\bibfnamefont{M.}~\bibnamefont{{Hannam}}},
  \bibinfo{journal}{\prd} \textbf{\bibinfo{volume}{91}}, \bibinfo{eid}{024043}
  (\bibinfo{year}{2015}), \eprint{1408.1810}.

\bibitem[{\citenamefont{{Buonanno} and {Damour}}(1999)}]{Buonanno:1999}
\bibinfo{author}{\bibfnamefont{A.}~\bibnamefont{{Buonanno}}} \bibnamefont{and}
  \bibinfo{author}{\bibfnamefont{T.}~\bibnamefont{{Damour}}},
  \bibinfo{journal}{\prd} \textbf{\bibinfo{volume}{59}}, \bibinfo{eid}{084006}
  (\bibinfo{year}{1999}), \eprint{gr-qc/9811091}.

\bibitem[{\citenamefont{{Damour} et~al.}(1998)\citenamefont{{Damour}, {Iyer},
  and {Sathyaprakash}}}]{Damour:1998}
\bibinfo{author}{\bibfnamefont{T.}~\bibnamefont{{Damour}}},
  \bibinfo{author}{\bibfnamefont{B.~R.} \bibnamefont{{Iyer}}},
  \bibnamefont{and} \bibinfo{author}{\bibfnamefont{B.~S.}
  \bibnamefont{{Sathyaprakash}}}, \bibinfo{journal}{\prd}
  \textbf{\bibinfo{volume}{57}}, \bibinfo{pages}{885} (\bibinfo{year}{1998}),
  \eprint{gr-qc/9708034}.

\bibitem[{\citenamefont{{Damour} and {Nagar}}(2008)}]{Damour:2008}
\bibinfo{author}{\bibfnamefont{T.}~\bibnamefont{{Damour}}} \bibnamefont{and}
  \bibinfo{author}{\bibfnamefont{A.}~\bibnamefont{{Nagar}}},
  \bibinfo{journal}{\prd} \textbf{\bibinfo{volume}{77}}, \bibinfo{eid}{024043}
  (\bibinfo{year}{2008}), \eprint{0711.2628}.

\bibitem[{\citenamefont{{Nagar} et~al.}(2019)\citenamefont{{Nagar}, {Messina},
  {Kavanagh}, {Lukes-Gerakopoulos}, {Warburton}, {Bernuzzi}, and
  {Harms}}}]{Nagar:2019}
\bibinfo{author}{\bibfnamefont{A.}~\bibnamefont{{Nagar}}},
  \bibinfo{author}{\bibfnamefont{F.}~\bibnamefont{{Messina}}},
  \bibinfo{author}{\bibfnamefont{C.}~\bibnamefont{{Kavanagh}}},
  \bibinfo{author}{\bibfnamefont{G.}~\bibnamefont{{Lukes-Gerakopoulos}}},
  \bibinfo{author}{\bibfnamefont{N.}~\bibnamefont{{Warburton}}},
  \bibinfo{author}{\bibfnamefont{S.}~\bibnamefont{{Bernuzzi}}},
  \bibnamefont{and} \bibinfo{author}{\bibfnamefont{E.}~\bibnamefont{{Harms}}},
  \bibinfo{journal}{\prd} \textbf{\bibinfo{volume}{100}}, \bibinfo{eid}{104056}
  (\bibinfo{year}{2019}), \eprint{1907.12233}.

\bibitem[{\citenamefont{Usman et~al.}(2016{\natexlab{b}})\citenamefont{Usman,
  Nitz, Harry, Biwer, Brown, Cabero, Capano, Canton, Dent, Fairhurst
  et~al.}}]{usman_pycbc_2016}
\bibinfo{author}{\bibfnamefont{S.~A.} \bibnamefont{Usman}},
  \bibinfo{author}{\bibfnamefont{A.~H.} \bibnamefont{Nitz}},
  \bibinfo{author}{\bibfnamefont{I.~W.} \bibnamefont{Harry}},
  \bibinfo{author}{\bibfnamefont{C.~M.} \bibnamefont{Biwer}},
  \bibinfo{author}{\bibfnamefont{D.~A.} \bibnamefont{Brown}},
  \bibinfo{author}{\bibfnamefont{M.}~\bibnamefont{Cabero}},
  \bibinfo{author}{\bibfnamefont{C.~D.} \bibnamefont{Capano}},
  \bibinfo{author}{\bibfnamefont{T.~D.} \bibnamefont{Canton}},
  \bibinfo{author}{\bibfnamefont{T.}~\bibnamefont{Dent}},
  \bibinfo{author}{\bibfnamefont{S.}~\bibnamefont{Fairhurst}},
  \bibnamefont{et~al.}, \bibinfo{journal}{Class. Quantum Grav.}
  \textbf{\bibinfo{volume}{33}}, \bibinfo{pages}{215004}
  (\bibinfo{year}{2016}{\natexlab{b}}), ISSN \bibinfo{issn}{0264-9381},
  \bibinfo{note}{publisher: IOP Publishing},
  \urlprefix\url{https://doi.org/10.1088/0264-9381/33/21/215004}.

\bibitem[{\citenamefont{and J~Aasi et~al.}(2015)\citenamefont{and J~Aasi,
  Abbott, Abbott, Abbott, Abernathy, Ackley, Adams, Adams, Addesso, Adhikari
  et~al.}}]{aligo2015}
\bibinfo{author}{\bibnamefont{and J~Aasi}},
  \bibinfo{author}{\bibfnamefont{B.~P.} \bibnamefont{Abbott}},
  \bibinfo{author}{\bibfnamefont{R.}~\bibnamefont{Abbott}},
  \bibinfo{author}{\bibfnamefont{T.}~\bibnamefont{Abbott}},
  \bibinfo{author}{\bibfnamefont{M.~R.} \bibnamefont{Abernathy}},
  \bibinfo{author}{\bibfnamefont{K.}~\bibnamefont{Ackley}},
  \bibinfo{author}{\bibfnamefont{C.}~\bibnamefont{Adams}},
  \bibinfo{author}{\bibfnamefont{T.}~\bibnamefont{Adams}},
  \bibinfo{author}{\bibfnamefont{P.}~\bibnamefont{Addesso}},
  \bibinfo{author}{\bibfnamefont{R.~X.} \bibnamefont{Adhikari}},
  \bibnamefont{et~al.}, \bibinfo{journal}{Classical and Quantum Gravity}
  \textbf{\bibinfo{volume}{32}}, \bibinfo{pages}{074001}
  (\bibinfo{year}{2015}),
  \urlprefix\url{https://doi.org/10.1088%2F0264-9381%2F32%2F7%2F074001}.

\bibitem[{\citenamefont{Acernese et~al.}(2014)\citenamefont{Acernese, Agathos,
  Agatsuma, Aisa, Allemandou, Allocca, Amarni, Astone, Balestri, Ballardin
  et~al.}}]{avirgo2014}
\bibinfo{author}{\bibfnamefont{F.}~\bibnamefont{Acernese}},
  \bibinfo{author}{\bibfnamefont{M.}~\bibnamefont{Agathos}},
  \bibinfo{author}{\bibfnamefont{K.}~\bibnamefont{Agatsuma}},
  \bibinfo{author}{\bibfnamefont{D.}~\bibnamefont{Aisa}},
  \bibinfo{author}{\bibfnamefont{N.}~\bibnamefont{Allemandou}},
  \bibinfo{author}{\bibfnamefont{A.}~\bibnamefont{Allocca}},
  \bibinfo{author}{\bibfnamefont{J.}~\bibnamefont{Amarni}},
  \bibinfo{author}{\bibfnamefont{P.}~\bibnamefont{Astone}},
  \bibinfo{author}{\bibfnamefont{G.}~\bibnamefont{Balestri}},
  \bibinfo{author}{\bibfnamefont{G.}~\bibnamefont{Ballardin}},
  \bibnamefont{et~al.}, \bibinfo{journal}{Classical and Quantum Gravity}
  \textbf{\bibinfo{volume}{32}}, \bibinfo{pages}{024001}
  (\bibinfo{year}{2014}),
  \urlprefix\url{https://doi.org/10.1088%2F0264-9381%2F32%2F2%2F024001}.

\bibitem[{\citenamefont{{Ashton} et~al.}(2019)\citenamefont{{Ashton},
  {H{\"u}bner}, {Lasky}, {Talbot}, {Ackley}, {Biscoveanu}, {Chu}, {Divakarla},
  {Easter}, {Goncharov} et~al.}}]{bilby}
\bibinfo{author}{\bibfnamefont{G.}~\bibnamefont{{Ashton}}},
  \bibinfo{author}{\bibfnamefont{M.}~\bibnamefont{{H{\"u}bner}}},
  \bibinfo{author}{\bibfnamefont{P.~D.} \bibnamefont{{Lasky}}},
  \bibinfo{author}{\bibfnamefont{C.}~\bibnamefont{{Talbot}}},
  \bibinfo{author}{\bibfnamefont{K.}~\bibnamefont{{Ackley}}},
  \bibinfo{author}{\bibfnamefont{S.}~\bibnamefont{{Biscoveanu}}},
  \bibinfo{author}{\bibfnamefont{Q.}~\bibnamefont{{Chu}}},
  \bibinfo{author}{\bibfnamefont{A.}~\bibnamefont{{Divakarla}}},
  \bibinfo{author}{\bibfnamefont{P.~J.} \bibnamefont{{Easter}}},
  \bibinfo{author}{\bibfnamefont{B.}~\bibnamefont{{Goncharov}}},
  \bibnamefont{et~al.}, \bibinfo{journal}{\apjs}
  \textbf{\bibinfo{volume}{241}}, \bibinfo{eid}{27} (\bibinfo{year}{2019}),
  \eprint{1811.02042}.

\bibitem[{\citenamefont{Howard and {others}}(2018)}]{howard_fastai_2018}
\bibinfo{author}{\bibfnamefont{J.}~\bibnamefont{Howard}} \bibnamefont{and}
  \bibinfo{author}{\bibnamefont{{others}}}, \emph{\bibinfo{title}{fastai}}
  (\bibinfo{publisher}{GitHub}, \bibinfo{year}{2018}),
  \urlprefix\url{https://github.com/fastai/fastai}.

\bibitem[{\citenamefont{Wang et~al.}(2016)\citenamefont{Wang, Yan, and
  Oates}}]{wang_time_2016}
\bibinfo{author}{\bibfnamefont{Z.}~\bibnamefont{Wang}},
  \bibinfo{author}{\bibfnamefont{W.}~\bibnamefont{Yan}}, \bibnamefont{and}
  \bibinfo{author}{\bibfnamefont{T.}~\bibnamefont{Oates}},
  \emph{\bibinfo{title}{Time {Series} {Classification} from {Scratch} with
  {Deep} {Neural} {Networks}: {A} {Strong} {Baseline}}} (\bibinfo{year}{2016}),
  \bibinfo{note}{1611.06455}.

\bibitem[{\citenamefont{He et~al.}(2015)\citenamefont{He, Zhang, Ren, and
  Sun}}]{he_deep_2015}
\bibinfo{author}{\bibfnamefont{K.}~\bibnamefont{He}},
  \bibinfo{author}{\bibfnamefont{X.}~\bibnamefont{Zhang}},
  \bibinfo{author}{\bibfnamefont{S.}~\bibnamefont{Ren}}, \bibnamefont{and}
  \bibinfo{author}{\bibfnamefont{J.}~\bibnamefont{Sun}},
  \bibinfo{journal}{CoRR} \textbf{\bibinfo{volume}{abs/1512.03385}}
  (\bibinfo{year}{2015}), \bibinfo{note}{1512.03385},
  \urlprefix\url{http://arxiv.org/abs/1512.03385}.

\bibitem[{\citenamefont{He et~al.}(2018)\citenamefont{He, Zhang, Zhang, Zhang,
  Xie, and Li}}]{he_bag_2018}
\bibinfo{author}{\bibfnamefont{T.}~\bibnamefont{He}},
  \bibinfo{author}{\bibfnamefont{Z.}~\bibnamefont{Zhang}},
  \bibinfo{author}{\bibfnamefont{H.}~\bibnamefont{Zhang}},
  \bibinfo{author}{\bibfnamefont{Z.}~\bibnamefont{Zhang}},
  \bibinfo{author}{\bibfnamefont{J.}~\bibnamefont{Xie}}, \bibnamefont{and}
  \bibinfo{author}{\bibfnamefont{M.}~\bibnamefont{Li}}, \bibinfo{journal}{arXiv
  e-prints} p. \bibinfo{pages}{arXiv:1812.01187} (\bibinfo{year}{2018}),
  \bibinfo{note}{1812.01187}.

\bibitem[{\citenamefont{Fawaz et~al.}(2020)\citenamefont{Fawaz, Lucas,
  Forestier, Pelletier, Schmidt, Weber, Webb, Idoumghar, Muller, and
  Petitjean}}]{fawaz_inceptiontime_2020}
\bibinfo{author}{\bibfnamefont{H.~I.} \bibnamefont{Fawaz}},
  \bibinfo{author}{\bibfnamefont{B.}~\bibnamefont{Lucas}},
  \bibinfo{author}{\bibfnamefont{G.}~\bibnamefont{Forestier}},
  \bibinfo{author}{\bibfnamefont{C.}~\bibnamefont{Pelletier}},
  \bibinfo{author}{\bibfnamefont{D.~F.} \bibnamefont{Schmidt}},
  \bibinfo{author}{\bibfnamefont{J.}~\bibnamefont{Weber}},
  \bibinfo{author}{\bibfnamefont{G.~I.} \bibnamefont{Webb}},
  \bibinfo{author}{\bibfnamefont{L.}~\bibnamefont{Idoumghar}},
  \bibinfo{author}{\bibfnamefont{P.-A.} \bibnamefont{Muller}},
  \bibnamefont{and}
  \bibinfo{author}{\bibfnamefont{F.}~\bibnamefont{Petitjean}},
  \bibinfo{journal}{Data Mining and Knowledge Discovery}
  \textbf{\bibinfo{volume}{34}}, \bibinfo{pages}{1936} (\bibinfo{year}{2020}),
  \bibinfo{note}{publisher: Springer Science and Business Media LLC},
  \urlprefix\url{https://doi.org/10.1007%2Fs10618-020-00710-y}.

\bibitem[{\citenamefont{Bai et~al.}(2018)\citenamefont{Bai, Kolter, and
  Koltun}}]{bai_empirical_2018}
\bibinfo{author}{\bibfnamefont{S.}~\bibnamefont{Bai}},
  \bibinfo{author}{\bibfnamefont{J.~Z.} \bibnamefont{Kolter}},
  \bibnamefont{and} \bibinfo{author}{\bibfnamefont{V.}~\bibnamefont{Koltun}},
  \emph{\bibinfo{title}{An {Empirical} {Evaluation} of {Generic}
  {Convolutional} and {Recurrent} {Networks} for {Sequence} {Modeling}}}
  (\bibinfo{year}{2018}), \bibinfo{note}{1803.01271}.

\bibitem[{\citenamefont{Dempster et~al.}(2021)\citenamefont{Dempster, Schmidt,
  and Webb}}]{dempster_minirocket_2021}
\bibinfo{author}{\bibfnamefont{A.}~\bibnamefont{Dempster}},
  \bibinfo{author}{\bibfnamefont{D.~F.} \bibnamefont{Schmidt}},
  \bibnamefont{and} \bibinfo{author}{\bibfnamefont{G.~I.} \bibnamefont{Webb}},
  in \emph{\bibinfo{booktitle}{Proceedings of the 27th {ACM} {SIGKDD}
  {Conference} on {Knowledge} {Discovery}
  \&amp\${\textbackslash}mathsemicolon\$ {Data} {Mining}}}
  (\bibinfo{publisher}{ACM}, \bibinfo{year}{2021}),
  \urlprefix\url{https://doi.org/10.1145%2F3447548.3467231}.

\bibitem[{\citenamefont{Rumelhart and
  McClelland}(1987)}]{rumelhart_learning_1987}
\bibinfo{author}{\bibfnamefont{D.~E.} \bibnamefont{Rumelhart}}
  \bibnamefont{and} \bibinfo{author}{\bibfnamefont{J.~L.}
  \bibnamefont{McClelland}}, in \emph{\bibinfo{booktitle}{Parallel
  {Distributed} {Processing}: {Explorations} in the {Microstructure} of
  {Cognition}: {Foundations}}} (\bibinfo{year}{1987}), pp.
  \bibinfo{pages}{318--362}.

\bibitem[{\citenamefont{Hochreiter and
  Schmidhuber}(1997)}]{hochreiter_long_1997}
\bibinfo{author}{\bibfnamefont{S.}~\bibnamefont{Hochreiter}} \bibnamefont{and}
  \bibinfo{author}{\bibfnamefont{J.}~\bibnamefont{Schmidhuber}},
  \bibinfo{journal}{Neural Computation} \textbf{\bibinfo{volume}{9}},
  \bibinfo{pages}{1735} (\bibinfo{year}{1997}), ISSN \bibinfo{issn}{0899-7667},
  \bibinfo{note}{conference Name: Neural Computation}.

\bibitem[{\citenamefont{Chung et~al.}(2014)\citenamefont{Chung, Gulcehre, Cho,
  and Bengio}}]{chung_empirical_2014}
\bibinfo{author}{\bibfnamefont{J.}~\bibnamefont{Chung}},
  \bibinfo{author}{\bibfnamefont{C.}~\bibnamefont{Gulcehre}},
  \bibinfo{author}{\bibfnamefont{K.}~\bibnamefont{Cho}}, \bibnamefont{and}
  \bibinfo{author}{\bibfnamefont{Y.}~\bibnamefont{Bengio}},
  \emph{\bibinfo{title}{Empirical {Evaluation} of {Gated} {Recurrent} {Neural}
  {Networks} on {Sequence} {Modeling}}} (\bibinfo{year}{2014}),
  \urlprefix\url{https://arxiv.org/abs/1412.3555}.

\bibitem[{\citenamefont{Kingma and Ba}(2014)}]{kingma_adam_2014}
\bibinfo{author}{\bibfnamefont{D.~P.} \bibnamefont{Kingma}} \bibnamefont{and}
  \bibinfo{author}{\bibfnamefont{J.}~\bibnamefont{Ba}},
  \emph{\bibinfo{title}{Adam: {A} {Method} for {Stochastic} {Optimization}}}
  (\bibinfo{year}{2014}), \urlprefix\url{https://arxiv.org/abs/1412.6980}.

\bibitem[{\citenamefont{Pascanu et~al.}(2012)\citenamefont{Pascanu, Mikolov,
  and Bengio}}]{DBLP:journals/corr/abs-1211-5063}
\bibinfo{author}{\bibfnamefont{R.}~\bibnamefont{Pascanu}},
  \bibinfo{author}{\bibfnamefont{T.}~\bibnamefont{Mikolov}}, \bibnamefont{and}
  \bibinfo{author}{\bibfnamefont{Y.}~\bibnamefont{Bengio}},
  \bibinfo{journal}{CoRR} \textbf{\bibinfo{volume}{abs/1211.5063}}
  (\bibinfo{year}{2012}), \eprint{1211.5063},
  \urlprefix\url{http://arxiv.org/abs/1211.5063}.

\end{thebibliography}

\end{document}